\documentclass[12pt]{article}

\usepackage{amsmath,amssymb,amsfonts,amsthm}
\usepackage{mathtools,mathrsfs,times,bm,graphicx}
\usepackage{accents} %for thick bar
\usepackage{color}
\usepackage{enumitem}
\usepackage{hyperref}
\usepackage{natbib}

\usepackage[margin=1in]{geometry}
\graphicspath{ {fig/} }

\theoremstyle{plain}

\newtheorem*{Thm*}{Theorem}
\newtheorem*{Prop*}{Proposition}
\newtheorem*{Lemma*}{Lemma}
\theoremstyle{definition}

\newtheorem{Cond}{Condition}
\newtheorem*{Defn*}{Definition}
\newtheorem*{Cond*}{Condition}
\theoremstyle{remark}

\newtheorem*{Note*}{Note}
\newtheorem*{Example*}{Example}
\theoremstyle{plain}
\newtheorem{theorem}{Theorem}
\newtheorem{proposition}{Proposition}
\newtheorem{corollary}{Corollary}
\newtheorem{lemma}{Lemma}
\newtheorem*{theorem*}{Theorem}
\newtheorem*{proposition*}{Proposition}
\newtheorem*{lemma*}{Lemma}
\theoremstyle{definition}

\newtheorem*{definition*}{Definition}
\newtheorem*{condition*}{Condition}
\theoremstyle{remark}

\newtheorem*{note*}{Note}
\newtheorem*{example*}{Example}

\newcommand{\reflect}[1]{\mathsf{R}\left({#1}\right)}
\newcommand{\reflectwarp}[1]{\mathsf{R_{prec}}\left({#1}\right)}
\newcommand{\Phug}{\mathsf{P_{hug}}}
\newcommand{\Phop}{\mathsf{P_{hop}}}
\newcommand{\Pflip}{\mathsf{P_{flip}}}
\newcommand{\PhugR}{\mathsf{P_{hugR}}}
\newcommand{\PhugHess}{\mathsf{P_{hugH}}}
\newcommand{\pitil}{\widetilde{\pi}}
\newcommand{\gtil}{\widetilde{g}}
\newcommand{\Htil}{\widetilde{H}}
\newcommand{\xtil}{\widetilde{x}}
\newcommand{\vtil}{\widetilde{v}}
\newcommand{\ghat}{\widehat{g}}

\newcommand{\Hop}{{\emph{Hop} }}

\newcommand{\md}{\mathsf{d}}

\newcommand{\bigO}[1]{\ensuremath{\cO\left(#1\right)}}
\newcommand{\bg}{\ensuremath{\bm{g}}}
\newcommand{\Var}[1]{\mathsf{Var}\left[{#1}\right]}

\newcommand{\Reject}{\mathcal{R}}
\newcommand{\Accept}{\mathcal{A}}
\newcommand{\cO}{\ensuremath{\mathcal{O}}}
\newcommand{\cX}{\ensuremath{\mathcal{X}}}
\newcommand{\cR}{\ensuremath{\mathcal{R}}}
\newcommand{\cA}{\ensuremath{\mathcal{A}}}
\newcommand{\infixmax}{\ensuremath{\vee}}
\newcommand{\bsigma}{\ensuremath{\bm{\sigma}}}
\newcommand{\set}[1]{\ensuremath{\left\{{#1}\right\}}}
\newcommand{\Bernoulli}{\mathsf{Bernoulli}}
\newcommand{\simiid}{\ensuremath{\overset{\textrm{iid}}{\sim}}}
\newcommand{\Normal}{\mathsf{N}}
\newcommand{\diag}{\mathsf{diag}}
\newcommand{\GammaDist}{\mathsf{Gamma}}
\newcommand{\BetaDist}{\mathsf{Beta}}
\newcommand{\half}{\frac{1}{2}}
\newcommand{\bzero}{\ensuremath{\bm{0}}}
\newcommand{\bone}{\ensuremath{\bm{1}}}
\newcommand{\II}[1]{\mathbb{I}\left({#1}\right)} % indicator funcition
\newcommand{\EE}[1]{\mathbb{E}\left[{#1}\right]}
\newcommand{\Expect}[1]{\mathbb{E}\left[{#1}\right]}
\newcommand{\norm}[1]{\left|\left|{#1}\right|\right|}
\newcommand{\abspipes}[1]{\left|{#1}\right|}
\newcommand{\dd}[2][]{\ensuremath{\operatorname{\mathsf{d}}^{#1}\!{#2}}}

\newcommand{\Cov}[1]{\mathsf{Cov}\left[{#1}\right]}
\newcommand{\recip}[1]{\frac{1}{#1}}
\newcommand{\partfrac}[2]{\frac{\partial #1}{\partial #2}}
\newcommand{\dbyd}[2]{\frac{\dd{#1}}{\dd{#2}}}
\newcommand{\pbyp}[2]{\frac{\partial{#1}}{\partial{#2}}}
\renewcommand{\gets}{\longleftarrow}
\newcommand{\ie}{\emph{i.e.}}

\newcommand{\Zperp}{Z^\bot}
\newcommand{\Zpara}{Z^\parallel}
\newcommand{\zperp}{z^\bot}
\newcommand{\zpara}{z^\parallel}

\begin{document}
%\onehalfspacing

%%%%%%%%%%%%%%%%%%%%%%%%%%%%%%%%%%%%%%%%%%%%%%%%%%%%%%%%%%%%%%%%%%%%%%%%%%%%%%

%\title{Hug and Hop\\
%\large A discrete-time, non-reversible Markov chain Monte Carlo algorithm}
\title{Hug and Hop: a discrete-time, non-reversible Markov chain Monte Carlo algorithm}
\author{Matthew Ludkin\footnote{Department of Mathematics and
    Statistics, Lancaster University, UK.} \footnote{ORCiD: 0000-0002-3832-8322}\\
  Chris Sherlock$^*$ \footnote{ORCiD: 0000-0002-2429-3157}\\
}
\date{}
\maketitle

%\bigskip
\begin{abstract}
    We introduced the Hug and Hop Markov chain Monte Carlo algorithm for
  estimating expectations with respect to an intractable distribution.
  The algorithm alternates between two kernels: \emph{Hug} and
  \emph{Hop}. \emph{Hug} is a non-reversible kernel that repeatedly
  applies the bounce mechanism from the recently proposed
  Bouncy Particle Sampler to produce a proposal point far from the
  current position, yet on almost the same contour of the target
  density, leading to a high acceptance probability.  Hug is
  complemented by \emph{Hop}, which deliberately proposes jumps
  between contours and has an efficiency that degrades very slowly
  with increasing dimension.  There are many parallels between Hug and
  Hamiltonian Monte Carlo using a leapfrog integrator, including
  the order of the integration scheme, however
  Hug is also able to make use of local Hessian information without
  requiring implicit numerical integration steps, and its performance is not terminally affected by unbounded gradients of the log-posterior. We test Hug and Hop empirically on a variety of toy targets
  and real statistical models and find that it can, and often does,
  outperform Hamiltonian Monte Carlo.
\end{abstract}

\noindent%
  \vfill

%\newpage
\section{Introduction}
\label{sec:introduction}

Markov chain Monte Carlo (MCMC) algorithms approximate expectations
under an un-normalised target distribution of $\pi$ by simulating a
Markov chain with $\pi$ as its stationary distribution then computing
empirical averages over the simulated values of the
chain. Historically MCMC has been based on reversible Markov kernels
such as the Metropolis-Hastings kernel~\cite[]{Hastings1970} and
special cases and variations of this
\cite[e.g.][]{brooks2011handbook} since it is straightforward to
ensure that these target $\pi$. However, there has been much recent
interest in non-reversible kernels \citep[e.g.][]{BouchardCt2018,
  Fearnhead2018} which have the potential both in practice and in
theory to be more efficient than their reversible
counterparts \citep{Neal1998, Diaconis2000, Bierkens2015, Ma2018}. A
particular continuous-time non-reversible algorithm, the Bouncy
Particle Sampler \citep{PetersdeWith2012,BouchardCt2018} and
variations such as the coordinate sampler \citep{wu2018} and the
Discrete Bouncy Particle Sampler \citep{BouchardCt2018,sherlock2017discrete} and variations on both bouncy samplers
\citep{vanetti2017piecewise}, use occasional reflections of a
velocity in the hyperplane perpendicular to the current gradient to
eliminate (for continuous-time versions) or substantially reduce
(discrete-time versions) rejections of proposed moves.

%The BPS and its variants appear to be very efficient when examining
%individual components of the target, however certain functionals of
%the Markov chain, $X$, and in particular $\log \pi(X)$, mix much more
%slowly than the individual components, whether the algorithm acts in
%continuous-time \citep{bierkens2018} or discrete-time
%\citep{sherlock2017discrete}. We turn this fundamental problem with
%bouncy particle samplers to our advantage.

We introduce a novel, discrete-time, non-reversible sampling algorithm
which itself consists of two accept-reject MCMC kernels, applied in
alternation. Given a current value, the first kernel uses the bounce
mechanism of the bouncy particle samplers to evolve a skew-reversible approximation to
a flow with constant speed along a level set of $\pi$ so as to produce a proposal point
that is far from the current position, yet on almost the same
posterior contour, leading to a high acceptance probability; we denote
this contour-hugging kernel \emph{Hug}.
%We emphasise that Hug does not
%use a geodesic integrator such as SHAKE or RATTLE
%\citep[e.g.][]{leimkuhler2004simulating}; these integrators
%use implicit schemes, whereas hug is fully explicit and approximates
%both the dynamics and the constraint.

The second kernel complements the first by focusing on moving between
contours. It encourages the next state of the Markov chain to lie on a substantially 
different contour by proposing a new point from a distribution centered on the current point, with a 
high variance in the gradient direction, and a lower variance in
directions perpendicular to the gradient; we denote this kernel
\emph{Hop}, and the combination of the two \emph{Hug and Hop}. Pseudo-code for the full algorithm is given in Appendix \ref{app:algorithms}.

\subsection{Notation}\label{sec:notation}
Throughout the article the target is assumed to have a density of
$\pi$ with respect to Lebesgue measure.  The log-density is denoted by
$\ell(x)=\log \pi(x)$ and its gradient and Hessian are denoted by
$g(x)=\nabla \ell(x)$ and
$H(x)=[\partial^2 \ell/\partial x_i \partial x_j]$, while the unit
gradient vector is denoted by $\ghat(x)=g(x)/\norm{g(x)}$. For some
small $\epsilon>0$, when the negative Hessian is positive definite
with all eigenvalues above $\epsilon$, we write
$\Sigma(x) = - H(x)^{-1}$. Otherwise, we set
$\Sigma(x)=-L^\top \{1/(\abspipes{\Lambda}+\epsilon I_d)\} L$, where
$L^\top \Lambda L$ is the spectral decomposition of $H$ and
$|\Lambda|$ denotes the (diagonal) matrix whose elements are the
absolute values of the corresponding elements of
$\Lambda$. $\Sigma(x)$ can therefore be considered as a local
variance-covariance matrix with eigenvalues informed by the local
curvature along each principal component, whether this curvature is positive or
negative. Given $\Sigma(x)$, the matrix $A(x)$ always denotes a
$d \times d$ matrix square-root of $\Sigma(x)$; \ie,
$A(x)^\top A(x)=\Sigma(x)$.

For a matrix $M$, we use the shorthand $M^{-\top}=\left(M^{-1}\right)^\top$ and we refer to its induced $\ell^2$ norm as:
$\norm{M}_{I} = \sup_{x\in \mathbb{R}^d\setminus \{0\}}\norm{Mx}_2/\norm{x}_2$.

\section{The Hug and Hop kernels}\label{sec:hughop}

\subsection{The Hug kernel}\label{sec:hug}
Given a current velocity, $v$ and a gradient vector, $g$, at the current position,  
the Bouncy Particle Sampler reflects the velocity in the hyper-plane tangent to the gradient as follows:
\begin{equation}
 \label{eqn.def.reflect}
 ~\reflect{v;g}=v-2(v^\top \ghat)\ghat.
\end{equation}
 A single application of the Hug kernel
 repeatedly alternates straight-line movement using the current velocity with an application of this reflection move
to repeatedly `bounce' the current velocity off the hyperplane tangent to the
local gradient and hence keep the net movement in the gradient direction small. The proposal mechanism from a current sample point
$x=x_0$ samples an initial velocity, $v_0$, from a proposal distribution
$q$ which satisfies $q(v\mid x)=q(-v\mid x)$ but does not force initial velocity to be perpendicular to the current gradient. Given a time interval, $T$, and a number of bounces, $B$, both tuning parameters, the discretisation interval is set to $\delta=T/B$, and the Hug kernel repeats the
following $B$ times: firstly move to $x_b':=x_b + \delta v_b/2$, then
reflect the velocity in the gradient at $x_b'$:
$v_{b+1}=\reflect{v_b;g(x_b')}$, and finally move to
$x_{b+1}=x_b' + \delta v_{b+1}/2$. The steps below describe a single application of the kernel, $\Phug$.

%\begin{algorithm}
%  \caption{Hug.}
%  \label{alg:vanilla-hug}
%  \begin{algorithmic}
% \Require integration time, $T$; number of steps, $B$; initial
% value, $x_0$; symmetric proposal density $q(\cdot |x)$.
% \State $\delta\gets T/B$.
% \State Draw velocity $v_0\sim q(\cdot | x_0)$ .
% \For{$b = 0 , \ldots, B-1$}
% \State Move to $x_b' = x_b + \delta v_b/2$.
% \State Reflect: $v_{b+1}\gets~\reflect{v_{b};g(x_b')}$.
% \State Move to $x_{b+1} = x_b' + \delta v_{b+1}/2$.
% \EndFor
% \State Compute $\log \alpha = \left[\ell(x_B) +
% \log q(v_B|x_B)\right]-\left[\ell(x_0) + \log q(v_0
% |x_0)\right]$.
% \State Accept $x_B$ as the new position with probability $\alpha$; otherwise remain at $x_0$.
% \end{algorithmic}
%\end{algorithm}
\begin{itemize}
%  \label{alg:vanilla-hug}
\item[]  \texttt{Require}: integration time, $T$; \# steps, $B$; current
 value, $x$; symmetric proposal density $q(\cdot |x)$.
 \item[] $x_0\gets x$ and $\delta\gets T/B$.
 \item[] Draw velocity $v_0\sim q(\cdot | x_0)$ .
 \item[] \texttt{For} {$b = 0 , \ldots, B-1$}
   \begin{enumerate}
 \item[] Move to $x_b' = x_b + \delta v_b/2$.
 \item[] Reflect: $v_{b+1}\gets~\reflect{v_{b};g(x_b')}$.
 \item[] Move to $x_{b+1} = x_b' + \delta v_{b+1}/2$.
 \end{enumerate}
 \item[] Compute $\log r_{hug} = \ell(x_B) -\ell(x_0) +\log q(v_B|x_B)-\log q(v_0
 |x_0)$.
 \item[] With a probability of $\alpha_{hug}=1\wedge r_{hug}$, $x\gets x_B$; else $x\gets x$.
 \end{itemize}

$\Phug$, can be viewed as the composition of two reversible kernels
each of which preserves detailed balance with respect to the extended
target of $\pitil(x,v):=\pi(x)q(v\mid x)$.  Let $\PhugR$ be exactly as
$\Phug$, except that the proposed velocity is $-v_B$ rather than
$v_B$, and let $\Pflip:(x,v)\rightarrow (x,-v)$, so that
$\Phug=\Pflip\PhugR$.  Since $q$ is symmetric, $\Pflip$ preserves
$\pitil$.  To see that $\PhugR$ preserves $\pitil$, and hence so
does $\Phug$, we first consider the loop within $\PhugR$.  The
transformation involves a reflection of velocity, sandwiched between
two translations of position; each of these individual transformations
has a Jacobian of $1$ and so the Jacobian for the entire
transformation from $(x_0,v_0)$ to $(x_B,-v_B)$ is also $1$. Hence, if
$X$ is stationary, the joint density of $(x_B,-v_B)$ is equal to
$\pi(x_0)q(v_0\mid x_0)$.  Secondly, the loop is skew symmetric,
so that starting from $(x_B,-v_B)$ and iterating the loop $B$ times, then flipping the velocity would lead back to $(x_0,v_0)$, so, at stationarity, the joint density for the reverse move is $\pi(x_B)q(-v_B\mid x_B)=\pi(x_B)q(v_B\mid x_B)$. Hence the acceptance probability $\alpha$ in the Hug Algorithm leads to $\PhugR$ being reversible with respect to $\pi(x)q(v\mid x)$.

\subsection{Error analysis for Hug}
To show why the hug kernel is effective as an MCMC proposal mechanism,
consider the step from $x_b$ to $x_{b+1}$. Taylor expanding about the
bounce point $x_b'$, and noting that
$x_b' - x_b = \frac{\delta}{2} v_b$ and
$x_{b+1} - x_b' = \frac{\delta}{2} v_{b+1}$ gives:
\begin{align*}
 \ell(x_b) & = \ell(x_b') - \frac{\delta}{2} v_b^\top g(x_b') + \frac{\delta^2}{8} v_b^\top H(x_b^+) v_b, \\
 \ell(x_{b+1}) & = \ell(x_b') + \frac{\delta}{2} v_{b+1}^\top g(x_b') + \frac{\delta^2}{8} v_{b+1}^\top H(x_{b+1}^-) v_{b+1},
\end{align*}
where $x_b^+$ lies on the line between $x_b$ and $x_b'$, and
$x_{b+1}^-$ lies on the line between $x_b'$ and $x_{b+1}$. Now
$(v_b + v_{b+1})^\top g(x_b') = 2 v_b^\top g(x_b') - 2(v_b^\top
\ghat(x_b')) \norm{g(x_b')} = 0$, so:
\begin{equation}
 \label{eqn.Taylor.diff}
 \ell(x_{b+1})-\ell(x_b) = \frac{\delta^2}{8} \left[v_{b+1}^\top H(x_{b+1}^-) v_{b+1} - v_b^\top H(x_b^+) v_b\right].
\end{equation}

Integrating for a time $T= B\delta$ requires $T/\delta$ such steps and
might be supposed to lead to an error of $\bigO{\delta}$. However, due to the special structure of the path, if the Hessian is well behaved the full integration also
has an error of $\bigO{\delta^2}$. We require the following conditions to obtain the theorem that follows, which is proved in
Appendix~\ref{sec:hug-scaling-analysis}. %We choose to require conditions
%based on the induced $\ell^2$-norm of the Hessian and of changes in the
%Hessian.

\begin{Cond}[Lipshitz-continuous Hessian]\label{cond:lipshitz-hess}~ \\
 There exists some $\gamma > 0$, such that
 $\norm{H(y) - H(x)}_{I} \leq \gamma\norm{y-x}$ for all
 $x,y \in \mathbb{R}^d$.
\end{Cond}

\begin{Cond}[Bounded Hessian]~\label{cond:bounded-hess} \\
 There exists some $\beta > 0$, such that
 $\sup_{x \in\mathbb{R}^d} \norm{H(x)}_{I} \leq \beta < \infty$.
\end{Cond}

\begin{theorem}\label{thm:hug-total-error}
 Consider a target $\pi$ such that the Hessian $H(x)$ of
 $\ell(x) = \log\pi(x)$ satisfies Conditions~\ref{cond:lipshitz-hess}
 and ~\ref{cond:bounded-hess}. For a single iteration of Hug
  with initial velocity $v_0$,
 \[
 |\ell(x_B) - \ell(x_0)| \leq \frac{1}{8}\delta^2\norm{v_0}^2 (2\beta + \gamma D),
\]
where $D = \norm{v_0T}$ is the total distance travelled in time $T$.
\end{theorem}

The larger $\beta$ and/or $\gamma$, the smaller $\delta$ must be. Potential consequences when Conditions 1 and/or 2 are not satisfied are illustrated in Appendix \ref{sec:stability-hug}. In practice, the size of $\delta$ that can be safely chosen is limited by the most extreme curvature on any surface of constant $\pi$ along which which large moves will be needed. 

The only velocity changes are reflections, so
$\norm{v_B}=\norm{v_0}$. Thus if $q$ is isotropic and independent of
$x$, rather than simply symmetric, then
$\alpha = 1 \wedge \exp[\ell(x_B)-\ell(x_0)] = \bigO{\delta^2}$. In
practice, for the standard version of $\Phug$ we choose a 
$q$ that is independent of $x$, and potential global anisotropy can be dealt with
by pre-conditioning, as we now discuss.

\subsection{Preconditioning of Hug}
\label{sec.Hug.precon}
Typically, preconditioning according to the overall shape of the
target can lead to large improvements in efficiency
\citep[e.g.][]{RR2001,SFR2010}. As in many other algorithms, such as
the random-walk Metropolis \citep{Hastings1970} or Metropolis-adjusted Langevin algorithm \citep{Besag1994,Roberts1998},
the shape of the proposal distribution should aim to mimic the shape
of the target and it might be preferable to employ an elliptically
symmetric proposal such as $V_b\mid x_b\sim \Normal(x_b,\Sigma)$,
where $\Sigma$ is some approximation to the variance matrix of $X$
under $\pi$.  The target $\tilde{x}=A^{-\top} x$,
where $A^\top A=\Sigma$, has $\mbox{Var}[\widetilde{X}]=I$, and a natural, isotropic proposal on this target is equivalent
to the elliptical proposal on the original target.
However, the bounce kernel also has a reflection move, and the
standard bounce dynamics, which have no \emph{a priori} understanding
of the target shape should be applied in the transformed,
approximately isotropic, space.  Since $\gtil(\xtil)=A g(x)$, this is
equivalent to applying the following reflection operator in the
original space \citep[][]{PGCP2017,sherlock2017discrete}:
\begin{equation}
 \label{eqn.def.reflect.warp}
 ~\reflectwarp{v;g}=v-2\frac{(v^\top g)}{g^\top\Sigma g}\Sigma g.
\end{equation}
The overall effect of preconditioning can be understood in terms of
Theorem~\ref{thm:hug-total-error} and Conditions
\ref{cond:lipshitz-hess} and~\ref{cond:bounded-hess} as effectively
reducing $\gamma$ and $\beta$ for a fixed $||v_0||$ and $T$, thus
allowing a larger step size, $\delta$.

The Hug proposal can also 
make explicit use of the Hessian during the velocity bounces, leading
to what is referred to in \citet{girolami2011riemann} as \emph{position-specific preconditioning}.  For each bounce point, $x'$, rather than
bouncing off the plane tangential to the gradient at $x'$, the kernel
$\PhugHess$ employs \eqref{eqn.def.reflect.warp}, but where
$\Sigma=\Sigma(x')=A(x')^\top A(x')$, where $A(x')$ is as defined in
Section~\ref{sec:notation}. Equivalently, just prior to each bounce, a
position-specific linear transformation is applied, the
reflection \eqref{eqn.def.reflect} is performed in the transformed
space, and then the linear transformation is reversed. Since the
particle's position has not changed during this process, neither has
$A(x')$. The algorithm is given in  Appendix~\ref{app:algorithms}.
 This kernel, $\PhugHess$, is also
skew-reversible and has a Jacobian of $1$.  The only
difference when compared to the vanilla Hug algorithm is the
reflection operation.  This also has a Jacobian of $1$ (it is a
reflection) and only uses information available
at $x'$.  Therefore, $\PhugHess$ is skew-reversible and
volume-preserving. Unlike for $\Phug$ where we usually choose
$q(v\mid x)$ to be independent of position, for $\PhugHess$, typically
$q(v\mid x)$ depends on $x$ through the Hessian at $x$.

Interestingly, a position-dependent transformation improves on the
$\mathcal{O}(\delta^2)$ error for a single step in
\eqref{eqn.Taylor.diff}; however, it is not possible to improve the
overall order of the algorithm. As with preconditioning, efficiency
gains arise from the effective reduction of $\beta$ and $\gamma$.
Proposition~\ref{prop.HugHessOne} is proved in Appendix~\ref{sec:hug-proveHessprop}.

\begin{proposition}
 \label{prop.HugHessOne}
 If $H(x)$ satisfies Condition~\ref{cond:lipshitz-hess},
 $|\ell(x_{b+1})-\ell(x_b)|\le
 \frac{\gamma\delta^3}{8}\left\{\norm{v_{b+1}}^3+\norm{v_b}^3\right\}$.
\end{proposition}

Contour-hugging alone will not explore the target well since, by
design, all points lie approximately on the same contour of the
target.  To ensure satisfactory exploration of the target, the
contour-hugging kernel is complemented by a contour-hopping kernel
which aims to propose points on different contours.

\subsection{The Hop kernel}\label{sec:hop}
We now describe the \emph{hop} kernel, which makes reversible moves
between contours by using gradient information to deliberately direct
most of the movement of a random-walk-style proposal either up or down
in the gradient direction.  For a given scaling, $\lambda_x$, of the
along-gradient component of the kernel, typically, the steeper the
gradient itself at $x$, the larger the resulting change in
log-posterior between the proposed value, $y$ and the current value,
$x$.  Motivated by the wish to control the magnitude of
$\ell(y)-\ell(x)$, when $\norm{g(x)}$ is large we decrease the overall
scaling in proportion to $\norm{g(x)}$ and use the proposal distribution:
\begin{equation}
 \label{eq:hop-prop}
 Y\mid X=x \sim \mathsf{MVN}\left(x, \frac{1}{\norm{g(x)}^2} B_x\right) ~\text{ with }~ B_x = \mu^2I + (\lambda^2 - \mu^2)\ghat(x) \ghat(x)^\top.
\end{equation}
Notice, $\ghat(x)^\top B_x\ghat(x)=\lambda^2$ and for any unit vector
$e\perp \ghat(x)$, $e^\top B_x e=\mu^2$, therefore, with respect to any
orthonormal basis that starts with $\ghat(x)$,
$\Var{Y \mid x} =
\mathsf{diag}(\lambda^2,\mu^2,\dots,\mu^2)/\norm{g(x)}^2$.  The portion
of the proposal perpendicular to $\ghat(x)$ is an isotropic Gaussian
with a scaling of $\mu/\norm{g(x)}$ and along the gradient line the proposal is
Gaussian with a scaling of $\lambda/\norm{g(x)}$.
Given this interpretation 
both $B_x^{-1}$ and $B_x^{1/2}$ have simple tractable forms (see Appendix \ref{app.add.Hop}) enabling straightforward simulation, and calculation of the acceptance probability in $\mathcal{O}\{\mathsf{dim}(x)\}$ operations.

The Metropolis-Hastings acceptance probability is
$\alpha_{hop}(x,y)=1\wedge r_{hop}(x,y)$, where:
\begin{equation}
  \log r_{hop}(x,y) = \ell(y) - \ell(x)+ \log \Normal\left(x; y, B_y/\norm{g(y)}^2\right)%\\
  %  &~~~~~
  - \log \Normal\left(y; x, B_x/\norm{g(x)}^2\right).
 \label{eq:hop-log-accept}
\end{equation}
If $\mu = 0$ then a proposed point $y$ will have an acceptance
probability of zero unless the gradient $\bg(y)$ is parallel to
$\bg(x)$.  Thus, in general, a strictly
positive value for $\mu$ is required.

If the scaling by $\norm{g(x)}^2$ were omitted, the Hop algorithm
would be a special case of the Directional Metropolis-Hastings
algorithm of \citet{mallik2017directional}; however, unlike the algorithm
in \citet{mallik2017directional}, the Hop algorithm is specifically
intended for jumping between contours. As we shall see in Theorem
\ref{thm:hop_prod} below, which is proved in
Appendix~\ref{app:hop_prod_proof}, and in the simulations in
Section~\ref{sec:sims}, the position-dependent scaling brings enormous
 and, perhaps, unexpected gains in efficiency for typical targets. In Theorem \ref{thm:hop_prod} all densities are with respect to the appropriate Lebesgue measure.

\begin{theorem}
  \label{thm:hop_prod}
  Consider a sequence of targets, $\left\{\pi^{(d)}\right\}_{d=1}^\infty$, with the following product density:
  $$
\pi^{(d)}\left(x^{(d)}\right)=\exp\left\{\sum_{i=1}^d\ell_1\left(x_i^{(d)}\right)\right\}.
    $$
  We assume that $\ell_1\in C^3$ with
  \begin{equation}
    \label{eqn.deriv.bounds}
    |\ell_1''(x)|\le T
    ~~~\mbox{and}~~~
|\ell_1'''(y)-\ell_1'''(x)|\le L|y-x|,
  \end{equation}
  for some $L,T<\infty$,
  and for a random variable $X$ with a density of $\exp\{\ell_1(x)\}$,
    \begin{equation}
      \label{eqn.moment.cond}
      \Expect{\ell_1'(X)^4}<\infty,
      ~~~
      \Expect{\ell_1''(X)^2}<\infty,
      ~~~\mbox{and}~~~
      \Expect{\ell_1'''(X)^2}<\infty.      
    \end{equation}
    The Hop algorithm is applied to target $\pi^{(d)}$ using scalings of $\lambda_d$ and $\mu_d=(\lambda_d \kappa)^{1/2}$, where, as $d\rightarrow \infty$,
    \begin{equation}
      \label{eqn.lambda.behave}
      \lambda_d\rightarrow \infty
      ~~~\mbox{and}~~~
      \lambda_d d^{-1/2}\rightarrow 0.
    \end{equation}
    Let $\alpha_{hop}^{(d)}(x^{(d)}, y^{(d)})$ be the corresponding
    acceptance probability as defined in and above \eqref{eq:hop-log-accept} and let
    $U\sim \Normal\left(-\frac{1}{2}\kappa,1\right)$.
    Then
    for a proposal $Y^{(d)}$ from a current point $X^{(d)}$, as $d\rightarrow \infty$
    \begin{equation}
      \label{eqn.acclargeDelta}
      \begin{bmatrix}
        \alpha^{(d)}_{hop}(X^{(d)},Y^{(d)})\\
        \frac{1}{\lambda_d}\left\{\log \pi^{(d)}\left(Y^{(d)}\right)-\log\pi^{(d)}\left(X^{(d)}\right)\right\}
      \end{bmatrix}
      \Longrightarrow
      \begin{bmatrix}
        1\wedge \exp(\kappa U)\\
        U
        \end{bmatrix}.
    \end{equation}
    In particular, therefore, 
 \begin{equation}
 \label{eqn.limiting.acc.rate}
 \lim_{d\rightarrow \infty}\mathbb{E}\left[\alpha^{(d)}_{hop}(X^{(d)},Y^{(d)})\right]
 \to
 2\Phi\left(-\frac{\kappa}{2}\right).
\end{equation}
\end{theorem}

Theorem \ref{thm:hop_prod} suggests that the parameterisation of Hop should be thought of in terms of $\lambda$, the scaling in the gradient direction, and $\kappa$, and that the acceptance rate should only depend on $\kappa$. Further, by \eqref{eqn.acclargeDelta}, $\lambda$ should be chosen as large as possible since the aim of the algorithm is to make large changes in $\log \pi$; however, once $\lambda=\mathcal{O}(d^{1/2})$, the asymptotics breakdown and we find in practice that the acceptance rate drops towards zero. This is demonstrated empirically in Figure \ref{fig:cauchit_explore} of Appendix \ref{app:cr-calc} for the 100-dimensional Cauchit regression example of Section \ref{sec:CR}. In practice, therefore, for a given $\kappa$ we recommend increasing $\lambda$ until the asymptotics have broken down but the acceptance rate has not yet dropped too close to $0$.

For fixed $\lambda_d$, the Theorem suggests optimising the natural objective function of expected squared change in $\log \pi$ which is proportional to
$\Expect{U^2\{1\wedge \exp(\kappa U)\}}$.  This leads to $\mu=\kappa=0$, which violates the assumptions made in and above \eqref{eqn.lambda.behave} as well as contradicting both the simulation study in Section \ref{sec.numInvHop} and common sense since $\mu=\kappa=0$ is only sensible on an isotropic target.  Figure \ref{fig:cauchit_explore} also shows that for fixed, moderately sized $\lambda$, as $\kappa\downarrow 0$ the acceptance rate is relatively flat, rather than increasing to $1$ as suggested by \eqref{eqn.limiting.acc.rate}, and choosing very small $\kappa$ is not, in fact, optimal. Thus, Theorem \ref{thm:hop_prod} cannot be used directly to obtain either an optimal setting for $\lambda$ or $\kappa$ but does provide the $(\lambda,\kappa)$ re-parameterisation and an heuristic for choosing $\lambda$.

Since $||g(x)||=\mathcal{O}(d^{1/2})$, the result requires that the overall scaling in the gradient direction be $o(1)$ and that the scaling should be $o(1/d^{1/4})$ in each direction perpendicular to the gradient. This should be contrasted with the standard scalings for the random walk Metropolis and Metropolis-adjusted Langevin algorithm of, respectively,  $\mathcal{O}(1/d^{1/2})$ and $\mathcal{O}(1/d^{1/6})$ \citep[e.g.][]{RR2001}. Unsurprisingly, since it uses gradient information, Hop is uniformly superior to the random walk Metropolis. It also supports larger jumps in the gradient direction than the Metropolis-adjusted Langevin algorithm. Hop is inferior to the latter algorithm in the $d-1$ directions perpendicular to the gradient; however, this fits with the purpose of Hop, which is to explore along the gradient rather than throughout the entire space.

As for the Hug algorithm in Section~\ref{sec.Hug.precon}, the efficiency of Hop
can be improved by global or position-specific preconditioning. For position-specific preconditioning, $\tilde x = A(x)^{-\top} x$ and $\tilde g(\tilde{x})=A(x) g(x)$; global preconditioning fixes $A(x)=A$ for all $x$. Details of the proposal and of the formula for the log-acceptance ratio are provided in Appendix \ref{app.add.Hop}.

\subsection{Numerical investigations of the Hop algorithm}
\label{sec.numInvHop}
We now investigate the performance of the Hop algorithm across a variety of toy targets and tunings. This demonstrates the robustness of the conclusions from Theorem
\ref{thm:hop_prod} to targets which do not strictly satisfy the conditions of the Theorem, in particular \eqref{eqn.deriv.bounds}, and informs the tuning advice to be given in Section \ref{sec:parameter-tuning}.
In practise, to avoid issues with small $||g||$, we use a multiplier of $1/(1\infixmax\norm{g}^2)$
 rather than $1/\norm{g}^2$ in the variance of the
 proposal of \eqref{eq:hop-prop}.

We consider a target density which is, for each component $i=1,\dots,d$,
proportional to the product of a centred logistic density with scale $\sigma_i$ and a
$\Normal(0, a^2\sigma_i^2)$ density. The Gaussian ensures
that the Hessian of the log target does not approach zero in the tails
of the distribution; the larger $a$ the smaller the contribution from
the Gaussian. We denote this density by:
\begin{equation}
 \label{eqn.lga}
 \pi_{\mathsf{LG}}(x; a, \underline{\sigma})
 \propto
 \prod_{i=1}^d \left[ \exp\left(-\frac{x_i}{2\sigma_i}\right) + \exp\left(\frac{x_i}{2\sigma_i}\right)\right]^{-2} \exp\left(-\frac{1}{2}\left(\frac{x_i}{a \sigma_i}\right)^2\right).
\end{equation}
We consider $a\in\{1,2,5\}$,
$d\in\{10,25,50,100,250\}$, $\lambda$ values between $1/8$ and $64$
and $\kappa$ values between $1/8$ and $4$. We choose different types
of target by changing the vector $\underline{\sigma}$: for \emph{i.i.d.} targets,
we set $\sigma_i=1$ for $i=1\to d$, whereas
for \emph{Linear} targets $\sigma_i = 1 + 9 \times (i-1)/(d-1)$.
In each combination of $a, d, \lambda, \kappa$ and target type,
we ran \emph{Hop} for $250,000$ iterations.

\begin{figure}[!ht]
  \centering \includegraphics[width=1.0\textwidth]{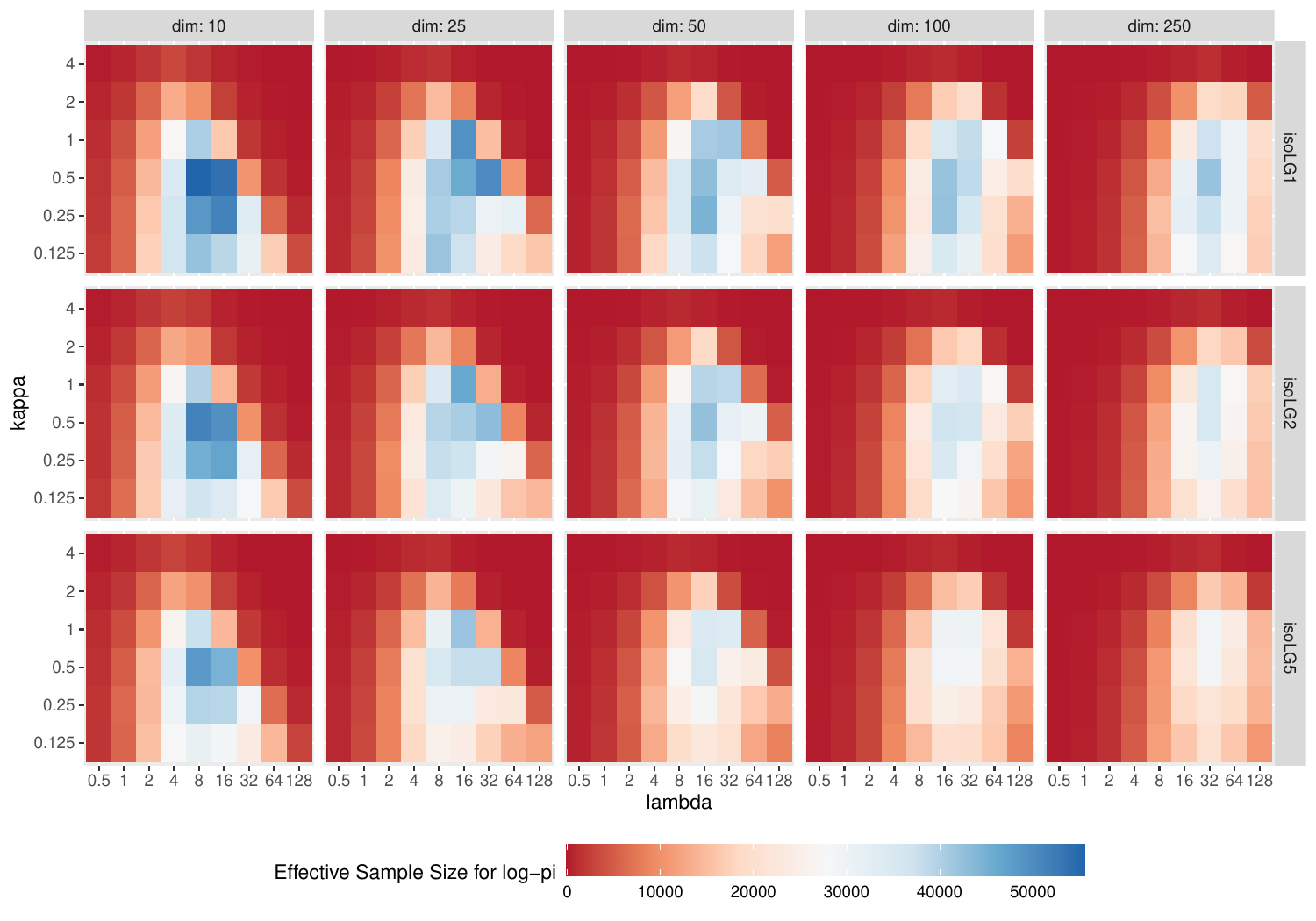}
  \caption{ Effective sample size of $\ell(X)$ under \emph{Hop} on a
    range of \emph{i.i.d.} targets (rows) and dimensions (columns). Within each
    cell, $\lambda$ and $\kappa$ are varied on an logarithmic scale
    (base 2).}
  \label{fig:hop-scaling-iso}
\end{figure}

\begin{figure}[!ht]
  \centering \includegraphics[width=1.0\textwidth]{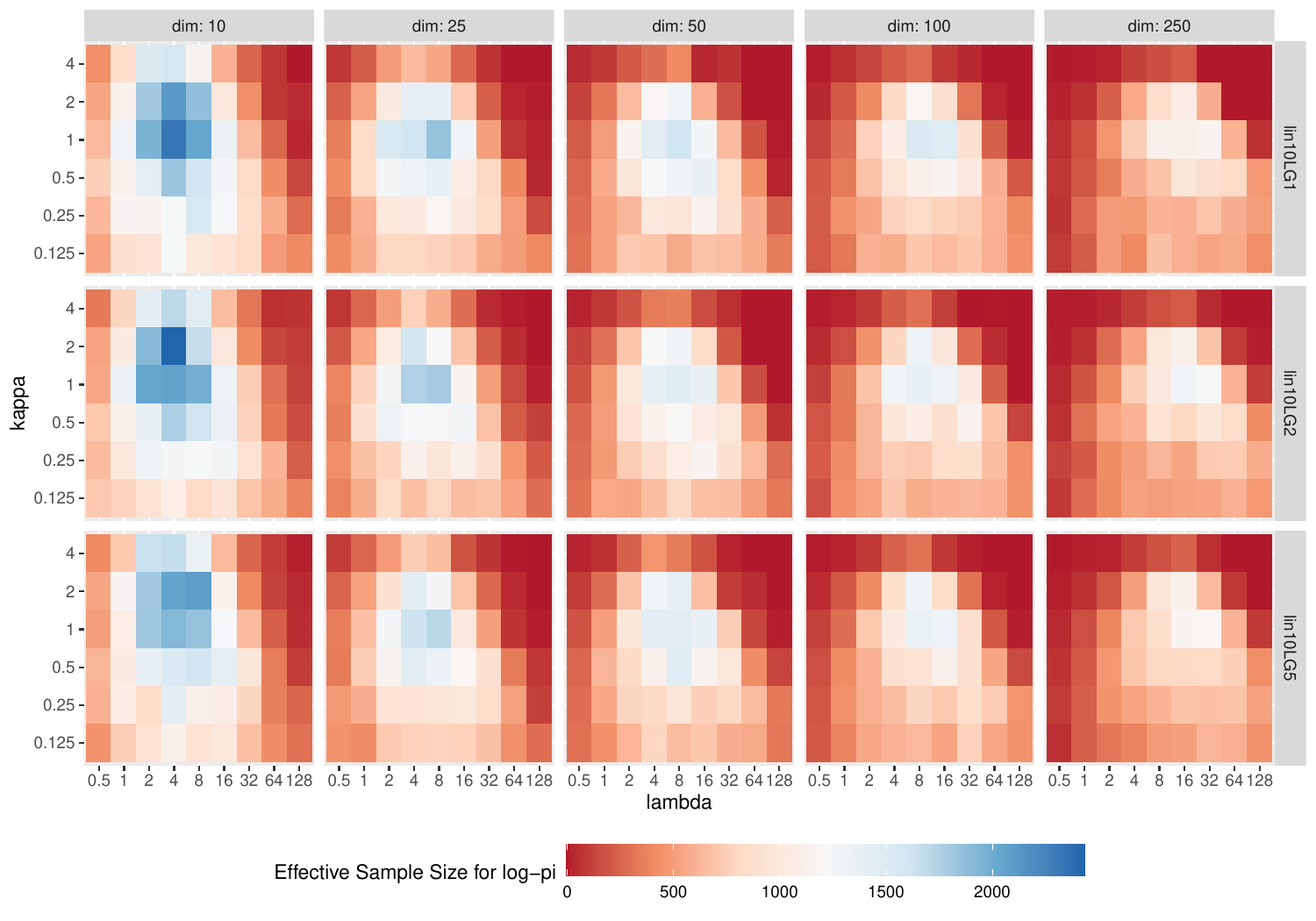}
  \caption{ Effective sample size of $\ell(X)$ under \emph{Hop} on a
    range of targets (rows) with scales increasing from $1$ to $10$ with 
    dimension index, and dimensions (columns). Within each
    cell, $\lambda$ and $\kappa$ are varied on an logarithmic scale
    (base 2).}
  \label{fig:hop-scaling-sqrt}
\end{figure}

\emph{Hop} is designed to move between contours of $\log \pi$.
The effective sample size of $\log \pi_{\mathsf{LG}}(X; 0, \sigma, a)$ as a function of the choice
of $a,d,\lambda, \kappa$ is shown in Figure~\ref{fig:hop-scaling-iso}
for \emph{i.i.d.} targets and in Figure~\ref{fig:hop-scaling-sqrt} for
\emph{Linear} targets. Firstly, whatever the target, the
optimal $\lambda$ increases with dimension just slightly slower than
in proportion to $d^{1/2}$. By contrast the optimal $\kappa$ is
remarkably stable across targets and dimension, lying between $0.25$
and $1$ for \emph{i.i.d.} targets and between $1$ and $2$ for \emph{linear} targets; recall that on a perfectly isotropic target detailed balance could be satisfied with $\kappa=0$ since the gradients at the current and proposed values would align.  For each combination of dimension and target, the
acceptance rates at the optimal $(\lambda,\kappa)$ values were between
0.1 and 0.46. Moreover, all plots show that the effects of $\lambda$
and $\kappa$ on performance are approximately orthogonal to each
other, backing up the reparameterisation from $(\lambda,\mu)$ to
$(\lambda,\kappa)$ suggested by Theorem~\ref{thm:hop_prod}.

Finally, \emph{Hop} efficiency degraded exceptionally slowly with
dimension in the \emph{i.i.d.} case: the ratio of the optimal effective sample size (ESS) with
$d=250$ to the optimal ESS with $d=10$ was $0.75$ for ISOLG1, $0.68$
for ISOLG2, and $0.61$ for ISOLG5.
In the case of \emph{Linear} targets, the corresponding ratios were
$0.50, 0.53, 0.56$ indicating a roughly 50\% reduction in efficiency when
dimension increases by a factor of 25.

\subsection{Parallels with Hamiltonian Monte Carlo}\label{sec:hmc-links}
 We now discuss the similarities
and differences between Hug and Hamiltonian Monte Carlo. Both algorithms augment the state space via a velocity $v_0$, which is typically drawn from a $\Normal(0,\Sigma)$ distribution.   Hug approximates the
movement for time $T$ along a level set of $\pi$ via a series of $B$ reflections, each
accounting for an integration time of $\delta=T/B$. Hamiltonian Monte Carlo approximates
the Hamiltonian dynamics of a particle moving in a potential of $-\ell$, which is equivalent movement along the level sets of total energy, via $L$ repeats of the Leapfrog
integrator \citep[e.g.][]{Neal2011}, each of which accounts for a time
$\delta=T/L$:
\begin{align*}
 v'=v_b+\frac{\delta}{2}\nabla \ell(x),~~
 x_{b+1}=x_b+\delta M^{-1}v',~~
 v_{b+1}=v'+\frac{\delta}{2}\nabla \ell(x_{b+1}),
\end{align*}
where $M=\Sigma^{-1}$ is the positive-definite mass matrix.  As with reflections,
each leapfrog step is skew reversible with a Jacobian of $1$, and so
the kernel targets $\pitil(x,v)=\pi(x)q(v\mid x)$ for exactly the same
reasons as Hug does. Also, as with {Hug} (Theorem
\ref{thm:hug-total-error}), the error in $\log \pitil$ after
integrating for a fixed time $T$ using steps of size $\delta$ is
$\mathcal{O}(\delta^2)$ \citep[e.g.][]{leimkuhler2004simulating}.

An appropriate choice of $M$ also allows for
global preconditioning of Hamiltonian Monte Carlo; however, any scheme that seeks to use local
Hessian information to set the mass matrix in the leapfrog step whilst
maintaining skew-reversibility must be \emph{implicit} and, hence,
much more time consuming: e.g. the middle step could become:
$x_{b+1}=x_b+\delta M^{-1}(\{x_b+x_{b+1}\}/2)v'$ \citep[see also][for
an implicit scheme which uses 3rd derivatives of
$\ell$]{girolami2011riemann}. This remains true if the alternative,
position-Verlet leaprog method is used.  The benefit of using local
Hessian information is demonstrated in Section~\ref{sec:sims}
\citep[see also][]{girolami2011riemann}.

The leapfrog step is symplectic and, as a consequence, if $\delta$ is
fixed and $\delta v_0$ is not too large given the curvature of
$\log \pi$, then as $T=L\delta$ increases the quantity
$|\ell(x_L)-\ell(x_0)|$ remains bounded
\citep[e.g.][]{leimkuhler2004simulating}. {Hug} is not symplectic;
nonetheless, we have found empirically that, as with the leapfrog
scheme, if $\delta v_0$ is not too large compared with the Hessian of
$\log \pi$, as $T=B\delta$ increases, $|\ell(x_B)-\ell(x_0)|$ remains
bounded; Figure~\ref{fig:hug-stability} in
Appendix~\ref{sec:stability-hug} demonstrates this empirically for
several different targets.
A final difference between the algorithms is in robustness to large gradient values, which we document next.

\subsection{Ergodicity and convergence}
\label{sec.ergANDconv}
On an isotropic target, neither Hug nor optimally tuned Hop is ergodic, as each algorithm is reducible. By the symmetry of the reflection operation, Hug remains on the same contour of $\log \pi$ forever. By contrast, since $g(y)$ is parallel to $g(x)$, Hop tuned with $\mu=0$ becomes a one-dimensional algorithm along a particular radial line; the same cancellation of large terms occurs, $\lambda$ can still be $o(d^{1/2})$, and the limiting acceptance rate is $1$. Though neither algorithm on its own is ergodic on such a target, Proposition \ref{prop.isotropic.ergodic} (proved in Appendix \ref{app.prove.ergodic.prop}) shows that the pair in tandem is. As mentioned in Section \ref{sec.numInvHop}, to avoid issues with very small gradients we replace the $1/||g(x)||^2$ term in \eqref{eq:hop-prop} by
$1/\left(1\vee ||g(x)||\right)^2$.

\begin{proposition}
  \label{prop.isotropic.ergodic}
  Let the distribution $\pi$ have a density with respect to Lebesgue measure on $\mathbb{R}^d$ of $f(||x||)$ for some $f:[0,\infty)\to(0,b)$, $b<\infty$. The Hug and Hop algorithm targeting $\pi$, with Hug using $v_0\sim \Normal(0,\tau^2 I_d)$, and with Hop using a proposal as in \eqref{eq:hop-prop} but with $||g(x)||^2$ replaced by $1\vee ||g(x)||^2$ is ergodic whether or not the $\mu$ scale parameter is zero.
\end{proposition}

Geometric ergodicity, convergence that is exponential in the number of iterations, is often deemed desirable. The two main classes of obstacles to geometric ergodicity are the existence of one or more regions of the space where the direction to the ``centre'' is difficult to discern, so the chain meanders (see, for example, Theorem 3.3 of \citep{MengTwee1996} or Theorem 4.3 of \citep{RobTwee1996MALA}), or where the acceptance rate can drop arbitrarily close to $0$ \citep[Proposition 5.1]{RobTwee1996RWM}. Local algorithms, such as the random-walk Metropolis, the Metropolis-adjusted Langevin algorithm and Hamiltonian Monte Carlo, as well as Hug and Hop, suffer from the former problem when the tails of the target decay slower than exponentially. The latter issue arises in gradient-based algorithms, including the Metropolis-adjusted Langevin algorithm \citep[Theorem 4.2]{RobTwee1996MALA}) and Hamiltonian Monte Carlo \citep[Theorem 2.2]{Livingetal2019} when the tails of the target are lighter than Gaussian, essentially because each leapfrog step includes two shifts of size $\mathcal{O}(||g(x)||)$, and $||g(x)||$ increases too quickly; however, this need not be an issue for Hug and Hop despite its use of gradients because Hug only depends on $g$ via $\ghat=g/||g||$, and in the presence of large gradients Hop \emph{reduces} the size of its jump proposals rather than increasing them.

 We formally show the robustness of Hop for the class of one-dimensional targets investigated in \citet{RobTwee1996MALA} and \citet{Livingetal2019}:
\begin{equation}
  \label{eqn.oneDtarget}
 \pi(x)\propto \exp\left(-\frac{1}{a}|x|^a\right),~~~x\in \mathbb{R}.
\end{equation}
For such targets, the random-walk Metropolis is known to be geometrically ergodic for $a \ge 1$ \citep[e.g.][Theorem 3.2]{MengTwee1996}, whereas the Metropolis-adjusted Langevin algorithm 
\citep[][Theorems 4.1, 4.2 and 4.3]{RobTwee1996MALA} and Hamiltonian Monte Carlo \citep[][Corollary 2.3]{Livingetal2019} are both geometrically ergodic only if either $1\le a < 2$, or, subject to an upper bound on the scale parameter, if $a=2$. Theorem \ref{thm.HopGE}, which is proved in Appendix \ref{app.GEHop}, shows that the Hop algorithm \emph{is} geometrically ergodic on light-tailed targets of the form \eqref{eqn.oneDtarget}. 

\begin{theorem}
  \label{thm.HopGE}
 A Metropolis-Hastings algorithm with a proposal as in \eqref{eq:hop-prop} but with $||g(x)||^2$ replaced by $1\vee ||g(x)||^2$ is geometrically ergodic on targets of the form \eqref{eqn.oneDtarget} provided $a\ge 1$.
\end{theorem}

We demonstrate empirically the convergence illustrated in Theorem~\ref{thm.HopGE} in a broader setting, using a target from
 \citet{sherlock2017discrete}:
\begin{align}\label{eq:expnormtarget}
\pi(x)\propto \exp\left(-\norm{x}_M^a/a\right),~~~x\in \mathbb{R}^d,
\end{align}
where
$\norm{x}^2_M = \sum_{i=1}^d x_i^2/\sigma_i^2$ and
$\bsigma = (1,\dots, d)$.
The mode of $\norm{X}_M$ when $X\sim\pi$ is $r^* = (d-1)^{1/a}$.

Firstly, we tuned Hamiltonian Monte Carlo and Hug and Hop to the main
body of the target in (\ref{eq:expnormtarget}) with $d=25$ and $a=4$.
This led, respectively, to $(T=2.0,L=4)$ and $(T=1.0, B=5,\lambda=10, \kappa=2)$. Using these tuning parameters, we repeated the following $N=50$ times for $\gamma\in \{1,1.5,2,2.5,3\}$, and each kernel $K \in \set{\textrm{Hamiltonian Monte Carlo}, \textrm{Hop}, \textrm{Hug and Hop}}$:
\begin{itemize}
\item Set $y= z/\norm{z}$, where $z \sim \Normal_{25}\left(0, I\right)$, so $y$ is uniform on the unit sphere.
\item Set the initial condition: $x_0 = \gamma r^* (\sigma_1 y_1, \ldots, \sigma_{d}y_{d})$ so that $||x||_M=\gamma r^*$.
\item Run kernel $K$ for 50,000 iterations and record the first time that $\norm{x}_M \leq r^*$.
\end{itemize}
At $\gamma=1$, the procedure
draws a point at the modal value of $\norm{X}_{M}$; as $\gamma$ increases, the initial value moves further into the tails of the target.  
The condition $\norm{x}_M \leq r^*$ is a proxy for convergence from
the tails to the posterior modal distance.  The results are given in
Figure~\ref{fig:hop_hmc_tail_iter} in Appendix \ref{sec.emp.Hop}.
% and Figure~\ref{fig:hop_hmc_tail_alpha}.  Notice that
Even when $\gamma=1.5$, Hamiltonian Monte Carlo fails to accept \emph{any} proposals
during 50,000 iterations and thus never converges by the above
condition.  In contrast, Hop converges for all values of
$\gamma$ considered, with only a slight increase in the time to
convergence.    Hop uses the gradient in two places: (i) to guide the variance
of the proposal which only uses the unit vector in the gradient
direction (and is thus not affected by the norm), and (ii) as the
scaling of the covariance matrix, which becomes small, not large, when the gradient
norm is large.

% \begin{figure}[ht!]
%   \centering
%   \includegraphics[width=0.8\textwidth]{tails_expnorm4_alpha}
%   \caption{Box plots of the acceptance rate during the 50,000
%     iterations for Hop and HMC on the target in Equation~(\ref{eq:expnormtarget}) with
%     $a=4$. The x-axis denotes starting multiplier $\gamma$.}
%   \label{fig:hop_hmc_tail_alpha}
% \end{figure}

% \red{
%   Notice
%   \[\begin{aligned}
%       \ell(x) &= -\norm{x}_M^a/a\\
%       (-a\ell(x))^{1/a} &= \norm{x}_M\\
%       \text{so}\qquad&\\
%       (-a\ell(x))^{1/a} = \norm{x}_M &\leq r^* = (d-1)^{1/a}\\
%       (-a\ell(x))^{1/a} &\leq (d-1)^{1/a}\\
%       -a\ell(x) &\leq (d-1)\\
%       \ell(x) &\geq \frac{1-d}{a}\\
%     \end{aligned}\]
% }

%\clearpage
\subsection{Parameter tuning}\label{sec:parameter-tuning}
\emph{Hug} and \emph{Hop} have different purposes, respectively to
move in $\mathcal{X}$ and to change $\log \pi$, and have separate
parameters, respectively $(T,B)$ and $(\lambda,\kappa)$. We recommend
tuning the pairs of parameters separately, each with the relevant goal
in mind.

For \emph{Hug}, as with HMC, $T$ should be large enough
that a reasonable distance is covered, but not so large that the
proposal dynamic is likely to perform a loop, making
$||x_B-x_0|| \ll ||Tv_0||$. Given $T$, $\delta$ should be
chosen so that the acceptance rate is bounded away from $0$ and
$1$. Empirical studies across a range of toy targets, dimensions and
integration times (see Appendix~\ref{sec.empirical.hug.eff}) suggest
setting $\delta$ so as to target an acceptance rate of between
$60\%$--$85\%$.

%For \emph{HugHess}, tuning is more complex as the term
%$q(v_B \mid x_B)/q(v_0\mid x_0)$ can dominate in the calculation of
%the acceptance probability when there are large changes in the Hessian
%between $x_0$ and $x_B$. Hence, $T$, also might need to be reduced to
%ensure that the acceptance probability is bounded away from $0$.

Tuning advice for \emph{Hop} derives from Theorem~\ref{thm:hop_prod}, the discussion thereafter and the simulation study of Section \ref{sec.numInvHop}.  With reasonable preconditioning, set
$\kappa \in [0.25, 1.0]$, perhaps a little larger if the preconditioning is poor. With small $\lambda$ this leads to the acceptance rate in \eqref{eqn.limiting.acc.rate}. For the chosen $\kappa$, increase $\lambda$ until the asymptotics no longer apply and the acceptance rate starts to decreases rapidly.  

\section{Simulation study}\label{sec:sims}
We now compare the Hug and Hop sampler to various
other algorithms on a range of target distributions in $d = 25$
dimensions.  We consider six classes of Model: (i) A 
Gaussian distribution with a diagonal co-variance matrix; (ii) a product
of a logistic density and a weak, regularising Gaussian density
$\pi_{\mathsf{LG}}(x; 5, \underline{\sigma})$, as defined in
\eqref{eqn.lga}; (iii) a product of a ``quartic'' and a weakly
regularising Gaussian:
\begin{equation}
  \label{eqn.qga}
  \pi_{\mathsf{QG}}(x; a, \underline{\sigma})
  \propto
  \prod_{i=1}^d\exp\left[-\frac{1}{4}\left(\frac{x_i}{\sigma_i}\right)^4\right] \exp\left[-\frac{1}{2}\left(\frac{x_i}{a \sigma_i}\right)^2\right]
\end{equation}
with $a=3$.  Models (iv) - (vi) are more exotic, each consists of a
$d=25$ dimensional target with dimensions $1$ and $2$ independent of
dimensions $3,\dots,d$, which themselves are independent, centred
Gaussians. The first two dimensions are: (iv) the Banana target of
\citep{sejdinovic2014kamh} with bananacity $\lambda=0.95$, (v) a
well-separated bimodal mixture of Gaussians and (vi) the Plus-Prism: a mixture of two Gaussians forming a ``+''-shaped target. Further details of these targets
 can be found in
Appendix~\ref{app:example-targ-calc}.

For each target, we consider two types of scaling across the
components: Isotropic scales, where the scale parameter of each
component is $1$, and Linear scales, where the scale parameter for component
$i$ is $1 + 24 \times (d-i)/(d-1)$. This yields $12$ targets.

The following MCMC algorithms were compared: the random walk
Metropolis (RWM), both vanilla and with Hessian-based proposal variance
\citep[e.g.][]{sejdinovic2014kamh}; the Metropolis-adjusted Langevin
algorithm (MALA) \citep[]{Besag1994,Roberts1998}; simplified manifold
MALA \citep[SMMALA,][]{girolami2011riemann}, which is MALA with
position-dependent preconditioning; Hamiltonian Monte Carlo; Hug
and Hop; Hug and Hop with both proposals using local Hessian information.

For each combination of target distribution and MCMC algorithm, to
allow a fair comparison, the algorithm was tuned over a grid of
parameter values
% to provide the best compromise between the per-second effective
% samples sizes (ESSs) of $X$ and $\ell=\log \pi(X)$. For each
% combination the two separate measures, minimum ESS over $X$
% components and ESS for $\ell$ were found
and the minimum effective sample size over all components of $X$ was found from a run
of $50,000$ iterations with the optimal parameter choice. Typically,
algorithm runs have a fixed computational budget or time limit, so
the total computational time for each run was also noted and
efficiency was measured in terms of the minimum effective sample size per second. To
compare the samplers, we consider values within each model relative
to the best for that model.

\begin{figure}[!ht]
 \centering \includegraphics[width=0.8\textwidth]{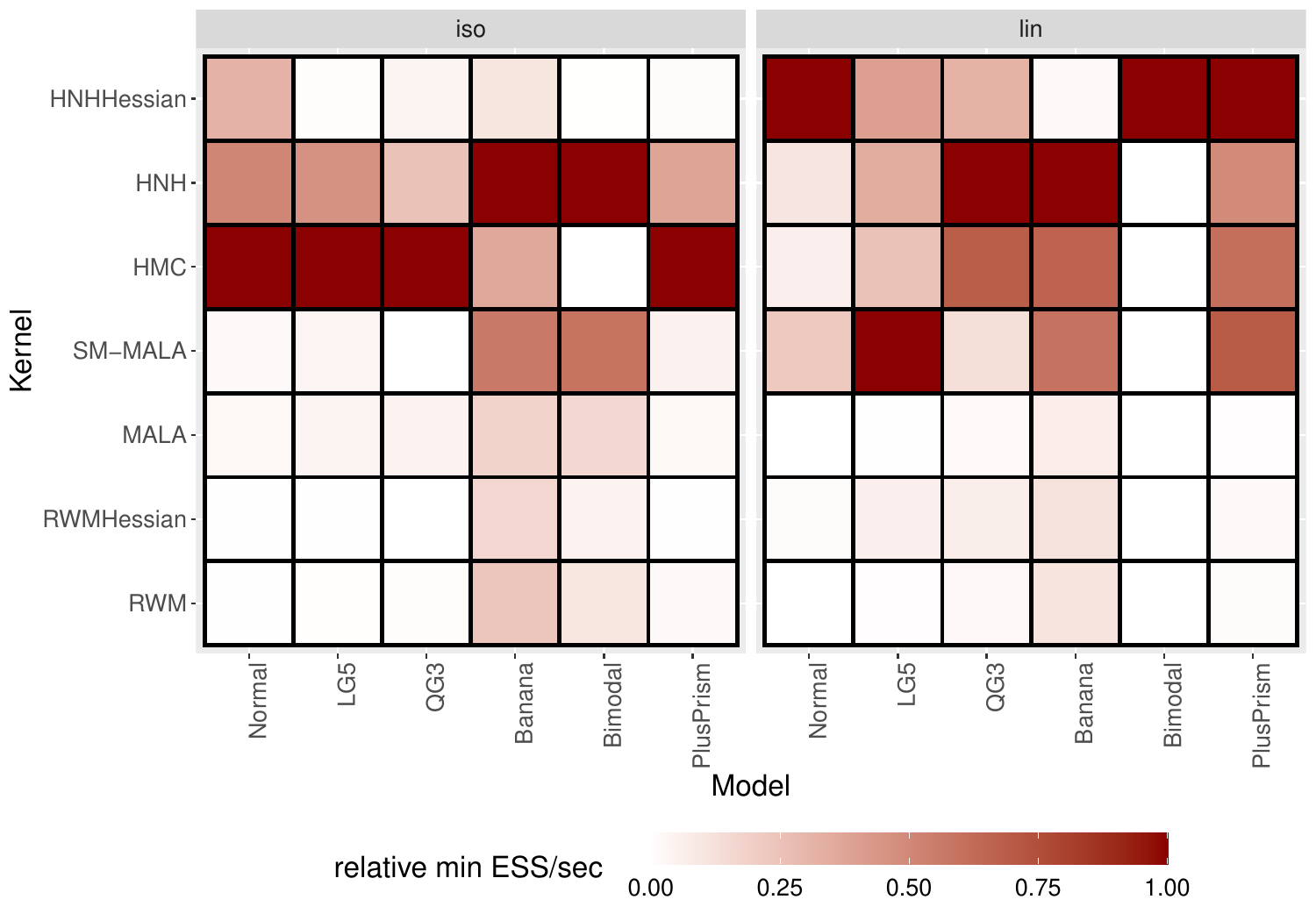}
 \caption{Minimum effective size (ESS) over the components of $x$ 
 per wall-clock second. Each sampler/target pair was run for 50,000 iterations on a 25- dimensional target. Values are relative to the most efficient algorithm within each model (columns). }
\label{fig:sim-ess-per-sec}
\end{figure}

The results, presented in Figure~\ref{fig:sim-ess-per-sec}, show that for unit targets Hug and Hop and Hamiltonian Monte Carlo are the most
efficient samplers, and for linear targets Hug and Hop using
position-dependent preconditioning is most efficient. The only
exception is the linear banana, where the standard Hug and Hop and Hamiltonian Monte Carlo are
more efficient than Hug and Hop using position-dependent
preconditioning.

The superiority of Hug and Hop on the bimodal target is of particular interest, and so we compare it, Hamiltonian Monte Carlo and the No U-turn Sampler of \citet{HoGe2014}, all without preconditioning, in a bimodal target stretched so that movement between modes happens only rarely. We tuned to obtain the maximum frequency of flips from one mode to the other taking CPU cost into account. The target and example trace plots are provided in Appendix \ref{sec.BimodalDetail}, along with detailed results. In summary, in this experiment, Hug and Hop is about $1.5$ times as efficient as Hamiltonian Monte Carlo, which is over twice as efficient as the No U-turn Sampler.

% in terms of ESS per second Hug and Hop (for the Unit targets) and
% Hug and Hop using Hessian information (for the {Linear} targets) are
% the most efficient, in terms of the minimum ESS of any component,
% with HMC (and sometimes SMMALA) in the next tier.  In contrast for
% the ESS of $\ell$, except for a bimodal target and the Gaussian
% targets, HMC (and sometimes SMMALA) performs better than Hug and
% Hop.

In general targets, Hessian calculations have a cost of $\mathcal{O}(d^2)$.
As dimension increases, use of position-dependent
conditioning would only remain of benefit if the eccentricity of the
contours also increased sufficiently quickly, or the
position-dependent conditioning was cheap to compute.

%\clearpage
\section{Statistical models}\label{sec:data}
In this section the utility of {Hug and Hop} is demonstrated and compared against Hamiltonian Monte Carlo on
some some real-world models, using simulated
data:
$30$- and $100$-dimensional Cauchit regression models;
an item-response, or Rash, model with $20$ tests and $100$ subjects;
% a $256$-dimensional probit regression model for binary spatial data,
and a 1002-dimensional stochastic volatility model. 
In the first two examples, we also test the No U-turn Sampler of \citet{HoGe2014}, where the recursive tree building leads to additional computational expense, so performance is measured in effective samples per CPU second. For the other example performance is evaluated via effective samples per gradient evaluation since gradient evaluations are by far the most computationally expensive operations performed each iteration. For fairness of comparison and to enable verification of our tuning advice, algorithms were tuned for maximum efficiency across a grid of parameter values as in Section \ref{sec:sims}.

\subsection{Cauchit regression}\label{sec:CR}
For data consisting of
binary responses with covariate information, the logistic or probit link functions are popular
choices for the Bernoulli GLM, but these link functions are not robust to outliers where the
linear predictor is large in absolute value, indicating the outcome is
almost certain, but the linear predictor is wrong
\citep{Koenker2009}. Such a situation may arise from errors in the
data-recording process, for example. The ``Cauchit'' link function is more tolerant
of such outliers. The model supposes that the $i$th binary response,
$Y_i$ is related to the vector of $M$ predictors for the response,
$x_i$, through some unknown parameters, $\beta$, as follows:
\begin{equation}
 \label{eq:CR}
 \begin{aligned}
 Y_i & \stackrel{\mathrm{indep.}}{{\sim}} \Bernoulli(p_i), \text{ for } i = 1 , \ldots, N \\
 g(p_i) & \simiid \beta^\top x_i,~\mbox{where}~g({u}) = \tan(\pi (u - 1/2)), \\
 \beta_j & \simiid \Normal\left(0, 1/\tau\right), \text{ for } i = 1 , \ldots, M.
\end{aligned}
\end{equation}

We simulated $M=30$ coefficients, $\beta_1,\dots,\beta_M$, from \eqref{eq:CR} with $\tau=1$, and $N=1000$ data points, $Y_i$, $i=1 , \ldots, N$, using predictors ($x_{i,1} , \ldots, x_{i,M}$), each of which was
independently drawn from a $\Normal\left(0,1\right)$ distribution. We compared the three algorithms
%Hug and Hop with Hamiltonian Monte Carlo and the No U-turn Sampler of \cite{HoGe2014}
on a posterior from \eqref{eq:CR} with $\tau=1$, starting each algorithm at the true parameter value and running for 50,000 iterations. We then repeated the simulation and analysis but with $M=100$. For extra robustness, both Hamiltonian Monte Carlo and Hug use jittering similar to that in \citet{Neal2011}: at each iteration, simulate $T^*\sim \mathsf{Unif}[0.8T,1.2T]$, then, respectively set $\epsilon=T^*/L$ or $\delta=T^*/B$.

With $M=30/100$, the optimal tunings were: $\epsilon=0.11/0.08$ for the {No U-turn Sampler}, which led to a mean number of leapfrogs per iteration of $\approx 7.08/14.91$; $(T,L)=(0.5,5)/(0.8,14)$ for {Hamiltonian Monte Carlo} and $(T,B,\lambda,\kappa)=(0.7,6,12,0.5)/(0.8,11,12,0.5)$ for Hug and Hop. Table \ref{tab:crm-ess} provides the acceptance rates and efficiencies at these tunings. For $M=30$, {Hug and Hop} is slightly more efficient than {Hamiltonian Monte Carlo}, whereas with $M=100$ {Hamiltonian Monte Carlo} is around $50\%$ more efficient. In each case, both are more efficient than the {The No U-turn Sampler}.

\begin{table}[ht]
  \centering
  \begin{tabular}{l|rrrrrr}
    Kernel                       & HMC    & Hug and Hop  & NUTS \\
    \hline\hline
    $\alpha$ ($M=30$)            &  71.6 & 74.6, 33.5 & 88.8\\
    Efficiency ($M=30$)   &  4630  & 5108   &   3248\\
\hline
    $\alpha$ ($M=100$)            &  86.1  & 72.4, 36.3&94.8 \\
    Efficiency ($M=100$)   &  1565  & 952   &  853 \\
  \end{tabular}
  % M=30
  % HMC  T=.5,L=5, ESS=13421, CPU=2899 (T=.5,L=6 also good)
  % HnH  T=.7,B=6, ESS=17408, CPU=3621 (T=0.6,B=6, ESS=17068, CPU=3414).
  % NUTS eps=0.11, ESS=12732, CPU=7840, lbar=7.08 - pretend CPU=7840/2
  % M=100
  % HMC  T=.8,L=14, ESS=33333, CPU=21294 (T=.7,L=12/13 also good)
  % H&H  T=.8,B=11, ESS=17119, CPU=17982 (T=.6, B=9 also good)
  % NUTS eps=.08, ESS=21002, CPU=49250, lbar=14.91 - pretend CPU=49250/2
  \caption{Acceptance rates and efficiency for the three algorithms on the Cauchit regression model with $N=1000$ and $M\in\{30,100\}$. Efficiency is measured in terms of the minimum over components of the effective sample size per 1000 CPU seconds.}
  \label{tab:crm-ess}
\end{table}

\subsection{Rasch model}\label{sec:rasch}

Consider a set of $N$ true or false questions answered by $M$ people.
Let $Y_{ij} = 1$ if person $i$ answered question $j$ correctly, and
$Y_{ij}=0$ otherwise. The Rasch model \citep[]{rasch1980} posits that
the $j$-th question has some latent difficulty $\beta_j$ and the
$i$-th person has a latent ability $\eta_i$ such that the probability
person $i$ is correct when answering question $i$ is given by
$P_{ij} =\Phi(\eta_i - \beta_j)$, where $\Phi$ is the distribution
function of a standard Gaussian.  Each answer $Y_{ij}$ is thus
considered as a Bernoulli outcome with probability $P_{ij}$.  Model identifiability can be ensured by arbitrarily fixing one of the parameters to $0$; however, there is no \emph{a priori} reason to believe this of any of the parameters. Our Bayesian analysis sidesteps the issue, keeping the exchangeability of the original model and ensures identifiability via the prior; we also do this as it  increases the correlation between the parameters, making the problem more challenging. The model is:
\begin{equation}
 \begin{aligned}
 \label{eq:Rasch}
 Y_{ij} & \stackrel{indep.}{{\sim}} \Bernoulli\left(\Phi(\eta_i - \beta_j)\right), \\
 \eta_i & \simiid \Normal\left(0, \tau^{-1}\right) \text{ for } i = 1 , \ldots, M,\\
 \beta_j & \simiid \Normal\left(0, \tau^{-1}\right) \text{ for } i = 1 , \ldots, N.
\end{aligned}
\end{equation}

 We simulated data from the model \eqref{eq:Rasch} with $M=100$ people, $N=20$ tests and $\tau=1$. For the subsequent
inference on $\theta=(\eta_1,\dots,\eta_M,\beta_1,\dots,\beta_N)$, the priors for $\eta_i$ and $\beta_j$ were as in the
model \eqref{eq:Rasch}.

A diagonal preconditioning matrix was used: $\Sigma=\mathsf{diag}(1,\dots,1,N/M,\dots,N/M)$, where the last $N=20$ elements are $N/M=0.2$. Hamiltonian Monte Carlo used a mass matrix of $M=\Sigma^{-1}$. Each sampler was run for 50,000 iterations, with jittering of $T$ applied as in Section \ref{sec:CR}.

The optimal tunings were $\epsilon=0.25$ for the No U-turn sampler, which led to a mean number of leapfrog steps per iteration of $7.01$, $(T,L)=(1.0,5)$ for Hamiltonian Monte Carlo,  and $(T,B,\lambda,\kappa)=(1.2,5,15,0.25)$ for Hug and Hop.
The results are given in Table~\ref{tab:irm-ess} and show that Hamiltonian Monte Carlo and
Hug and Hop have similar performance. Even though the former performs better
on $\eta$, Hug and Hop performs better on the worst mixing component, which is a component of $\beta$.

% \red{Recall that the algorithms were tuned to perform well on the
% components $\beta,\eta$. HMC ``focuses'' its effort on mixing on the
% components at the expense of $\log\pi$; whereas Hug and Hop separate
% these concerns: Hug focuses on components and hop focuses on
% $\log\pi$, therefore when tuning Hug and Hop to mix on the components,
% the mixing in $\log\pi$ suffers much less.}

\begin{table}[ht]
  \centering
  \begin{tabular}{lrrrrrr}
    Kernel                       & HMC    & Hug and Hop & NUTS\\
    \hline
    $\alpha$ (\%)                    &  79.2  & 86.6, 48.6 & 89.0 \\
    Efficiency($\eta$)    & 235 & 215   &150   \\
    Efficiency($\beta$)   &  105  & 121   &78   \\
    %% min ESS($\log\pi$)     & 3650    & 8703 & 6156         \\
    %% CPU time & 172.2 & 216.6 508.8
    %% but should divide NUTS by (just less than) 2 because
    %% I use two gradient calculations per leapfrog and
    %% better code would use 1 and some cunning storage.
  \end{tabular}
  \caption{The Rasch model with $M=100$ and $N=20$: acceptance rates and efficiencies for the three algorithms. Efficiency  both for the vector $\eta$ and the vector $\beta$ is measured in terms of the minimum over the vector of the effective sample size per CPU second.}
  \label{tab:irm-ess}
\end{table}

\subsection{Stochastic Volatility Model}\label{sec:stochvol}

Consider the following model for zero-centred data
$y=(y_0,\dots,y_{T-1})$ where the variance depends on a zero-mean,
Gaussian AR(1) process started from stationarity:
\[
\begin{aligned}
y_t &\sim \Normal\left(0, \frac{\exp(2x_t)}{\tau}\right),~t=0,\dots,T-1,\\
x_0 &= \frac{z_0}{\phi}~~~\mbox{and}~~~x_t = \rho x_{t-1} + z_t,~t=1,\dots,T-1,
\end{aligned}
\]
where $z_t\sim \Normal(0, 1),~t=0,\dots,T-1$ are iid. 
Parameter priors are $\tau~\sim \GammaDist(21, 5)$ and $(1+\rho)/2 \sim \BetaDist(20, 2)$, with $\phi=\sqrt{1-\rho^2}$.
Standard transformations ensure that all parameters have support on $(-\infty,\infty)$:
\[
\alpha = -\half \log(\tau)
, \quad
\beta = \half\left\{\log(1+\rho) - \log(1-\rho)\right\}.
\]
Appendix \ref{app:stochvol} provides the log posterior and its gradients
with respect to $z$, $\alpha$ and $\beta$.

We simulated data from the model using the parameters
$(\tau, \rho) = (4.0, 0.95)$ (see Figure~\ref{fig:sv-data} in Appendix
\ref{app:stochvol}). We then ran HMC with $T=3.0, L=35$ and Hug and Hop
with $T=3.75, B=35, \lambda = 10, \kappa=0.5$. In each case a diagonal
pre-conditioning matrix estimated from some initial runs was used.
For Hug and Hop, the Hop kernel was applied five times per iteration, rather than once, as this was found to improve the mixing at little extra computational cost.

Each sampler was initialised at a point well supported by the
posterior and run for 50,000 iterations. HMC uses 35 steps and thus 35 gradient
evaluations per iteration. For hug and hop, hug uses 35 evaluations and hop
(repeated five times) uses 5, giving a total of 40.
The acceptance rates were 87\% for HMC, 77\% for Hug, and 39\% for
Hop.  The worst mixing component was $\beta$, for which \emph{Hamiltonian Monte Carlo} is slightly more efficient than \emph{Hug and Hop}.

\begin{table}[ht]
  \centering
  \begin{tabular}{lrrrrrr}
    Kernel                   & HMC  & Hug and Hop \\
    \hline
    Acceptance rate $\alpha$ & 0.87 & 0.77, 0.39  \\
    Efficiency ($\alpha$)  & 2755 & 2410        \\
    Efficiency ($\beta$)   & 598  & 523         \\
    Efficiency ($Z$) & 1600 & 1400        \\
%    ESS($\log\pi$) per grad. & 213  & 186         \\
  \end{tabular}
  \caption{Stochastic Volatility model with $T=1000$: acceptance rates and efficiencies for $\alpha$, $\beta$ and the vector $Z$. Efficiency is measured in terms of the (minimum, for $Z$) effective sample size per 50000 gradient evaluations.}
  \label{tab:svm-ess}
\end{table}

% HMC ESS(alpha)/ngrads 2755.1560
% HMC ESS(beta)/ngrads   598.1501
% HMC min ESS(Z)/ngrads 1600.305
% HMC ESS(logpi)/ngrads  213.0816

% HNH ESS(alpha)/ngrads 2410.761
% HNH ESS(beta)/ngrads   523.3814
% HNH min ESS(Z)/ngrads 1400.267
% HNH ESS(logpi)/ngrads  186.4464

%We also considered the robustness of each algorithm: taking the tuning
%parameters for the above study, we started both algorithms at points
%in the tails of the target by setting $\beta=0$, drawing $z$ from it's
%prior and setting $\alpha=-1,-2 \to -6$ and running for 5000 iterations.

%No run of HMC ever accepted a single proposal, whereas Hop always had
%an acceptance rate above 50\% (and hug at least 20\%). The poor behaviour of HMC is due to
%the large norm of the gradient causing the leapfrog integrator to
%become unstable as discussed in Section~\ref{thm.HopGE}.

\section*{Acknowledgements}
Work by both ML and CS was supported by EPSRC grant EP/P033075/1.

\bibliographystyle{apalike}
\bibliography{ref}

\appendix

\section{Additional algorithm details}\label{app:algorithms}
One iteration of the full Hug-and-Hop algorithm with a symmetric proposal density $q(v |x)$ for Hug and with the Hop proposal robust to small gradient magnitudes proceeds as follows:

\begin{itemize}
\item[\texttt{Require}]  Hug time, $T$; Hug \# steps, $B$; Hop scale, $\lambda$; Hop ratio $\kappa$; current value, $x$.
\item[Hug:] $x_0\gets x$ and $\delta\gets T/B$.
\item[] Draw velocity $v_0\sim q(\cdot | x_0)$ .
\item[] \texttt{For} $b = 0 , \ldots, B-1$,
  \begin{itemize}
 \item[] Move to $x_b' = x_b + \delta v_b/2$.
 \item[] Reflect: $v_{b+1}\gets~\reflect{v_{b};g(x_b')}$.
 \item[] Move to $x_{b+1} = x_b' + \delta v_{b+1}/2$.
   \end{itemize}
\item[] \texttt{EndFor}
\item[] Compute
  $\log r_{hug} = \ell(x_B) - \ell(x_0) +\log q(v_B|x_B)- \log q(v_0 |x_0)$.
\item[] With a probability of $\alpha_{hug}=1\wedge r_{hug}$, $x\gets x_B$; otherwise $x\gets x$.
 \item[Hop:] Draw $y$ from $\Normal\left(x, \frac{1}{1\vee ||g(x)||^2 B_x}\right)$, where $B_x=\lambda \kappa I +(\lambda^2-\lambda \kappa)\ghat(x)\ghat(x)^\top$.
 \item[] Set $B_y=\lambda \kappa I +(\lambda^2-\lambda \kappa)\ghat(y)\ghat(y)^\top$ and compute
 \[\log r_{hop}=\ell(y) - \ell(x)+ \log \Normal\left(x; y, \frac{B_y}{1\vee\norm{g(y)}^2}\right) - \log \Normal\left(y; x, \frac{B_x}{1\vee \norm{g(x)}^2}\right).\]
 \item[] With a probability of $\alpha_{hop}=1\wedge r_{hop}$, $x\gets y$; otherwise $x\gets x$.
 \end{itemize}

In practice, in the above we often choose $q(v|x)$ to be $\Normal\left(v;0,I_d\right)$ so that the two $\log q$ terms in $r_{hug}$ cancel. When per-iteration jittering is used in Hug, the first line of Hug changes to $x_0\gets x$, $T_*\sim \mathsf{Unif}[0.8T,1.2T]$ and $\delta=T_*/B$. When pre-conditioning is used, the whole algorithm applies to the transformed posterior.

The hug algorithm with position dependent conditioning is given below. In practise we choose $q(v|x)$ to be 
$\Normal\left(v;0,\Sigma(x)\right)$.

\begin{itemize}
  \item[]  \texttt{Require}: integration time, $T$; \# steps, $B$; current value, $x$; position dependent scaling function $\Sigma(x)$.
    \item[] $x\gets x_0$ and $\delta\gets T/B$.
    \item[] Draw velocity $v_0\sim q(\cdot | x_0)$ .
    \item[] \texttt{For} {$b = 0 , \ldots, B-1$}
      \begin{itemize}
      \item[] Move to $x_b' = x_b + \delta v_b/2$.
      \item[] Reflect: $v_{b+1}\gets v_b - 2 \frac{v_b^\top g(x)}{g(x)^\top \Sigma(x) g(x)} \Sigma(x) g(x)$.
    \item[] Move to $x_{b+1} = x_b' + \delta v_{b+1}/2$.
\end{itemize}
\item[] \texttt{EndFor}
\item[] Compute
  $\log r_{hug} = \ell(x_B) - \ell(x_0) +\log q(v_B|x_B)- \log q(v_0 |x_0)$.
\item[] With a probability of $\alpha_{hug}=1\wedge r_{hug}$, $x\gets x_B$; otherwise $x\gets x$.
\end{itemize}

\section{Proofs of Theoretical results for Hug}
\subsection{Proof of Theorem~\ref{thm:hug-total-error}}
\label{sec:hug-scaling-analysis}
In this section we prove Theorem~\ref{thm:hug-total-error}.
\begin{proof}
  Firstly, write the difference in $\ell$ at $x_B$ and $x_0$ as a
 telescoping sum and apply Equation~(\ref{eqn.Taylor.diff}):
 \begin{align}
  \ell(x_B) - \ell(x_0)
  & = \sum_{b=1}^B \ell(x_b) - \ell(x_{b-1})
  & \text{[telescope]}\nonumber\\
  & = \frac{\delta^2}{8} \sum_{b=1}^B \left[v_b^\top H(x_b^-) v_b - v_{b-1}^\top H(x_{b-1}^+) v_{b-1}\right]
  & \text{[Equation~(\ref{eqn.Taylor.diff})]}\nonumber\\
  & =\frac{\delta^2}{8} \left(v_B^\top H(x_B^-) v_B - v_0^\top H(x_0^+) v_0
   + \sum_{b=1}^{B-1} \left[v_b^\top (H(x_b^-) - H(x_b^+)) v_b\right]\right)
  &~\label{eq:hug-scaling-split}
 \end{align}
 Recall that $x_b^-$ and $x_b^+$ lie on the line segment, namely the
 segment joining the bounce points $x_{b-1}'$ and $x_b'$.
 Furthermore, note that
 $x_b' = x_b + \delta v_b/2 = x_{b-1}' + \delta v_b$, therefore:
 \[
  \norm{x_b^+ - x_b^-} \leq \norm{x_b' - x_{b-1}'} = \delta\norm{v_i}.
\]
 This allows us to bound each term in the summation within
 (\ref{eq:hug-scaling-split}):
 \begin{align}
  |v_b^\top (H(x_b^-) - H(x_b^+)) v_b|
  &\leq \norm{v_b}\norm{(H(x_b^-) - H(x_b^+)) v_b}
  &\text{[Cauchy-Shwartz]} \nonumber\\
  &\leq \norm{v_b}^2 \norm{H(x_b^-) - H(x_b^+)}_{I}
  &\text{[Definition of induced norm]} \nonumber\\
  &\leq \gamma \norm{v_b}^2\norm{x_b^+ -x_b^-}
  &\text{[Condition~\ref{cond:lipshitz-hess}]} \nonumber\\
  &\leq \gamma \delta\norm{v_b}^3.\label{eq:hug-scaling-term2}
 \end{align}
 By Condition~\ref{cond:bounded-hess}, we can also bound the first
 difference in (\ref{eq:hug-scaling-split}):
 \begin{align}
  v_B^\top H(x_B^-) v_B - v_0^\top H(x_0^+) v_0
  \leq \beta (\norm{v_B}^2 + \norm{v_0}^2) = 2\beta \norm{v_0}^2,\label{eq:hug-scaling-term1}
 \end{align}
 where we use the fact $\norm{v_b} = \norm{v_{b-1}}$ since reflection
 preserves the norm.  Combining (\ref{eq:hug-scaling-term2}) and
 (\ref{eq:hug-scaling-term1}) in (\ref{eq:hug-scaling-split}) with the
 triangle inequality results in:
 \begin{align*}
  |\ell(x_B) - \ell(x_0)|
  &\leq \frac{\delta^2}{8} \left[
   \abspipes{v_B^\top H(x_B^-) v_B - v_0^\top H(x_0^+) v_0}
   + \abspipes{\sum_{b=1}^{B-1} v_b^\top (H(x_b^-) - H(x_b^+)) v_b}
   \right]\\
  & \leq \frac{\delta^2}{8} \left(2\beta\norm{v_0}^2
   + (B-1)\gamma\delta\norm{v_0}^3\right)\\
  & \leq \frac{\delta^2\norm{v_0}^2}{8} \left(2\beta
   + \gamma \norm{T v_0}\right),
 \end{align*}
 where the last line follows from $T=B\delta$.
\end{proof}

\subsection{Proof of Proposition~\ref{prop.HugHessOne}}
\label{sec:hug-proveHessprop}
\begin{proof}
  Without loss of generality, set $b=0$ and write $A$ for $A(x_0')$.
 Applying \eqref{eqn.Taylor.diff} but in the transformed space where
 the Hessian is $\Htil(x)=AH(\xtil)A^\top$, gives
 \begin{align*}
  |\ell(x_1)-\ell(x_0)|
  & = |\ell(\xtil_1)-\ell(\xtil_0)|\\
  & = \frac{\delta^2}{8} \abspipes{
   \vtil_1^\top AH(x_1')A^\top\vtil_1 - \vtil_0^\top AH(x_0')A^\top \vtil_0
   }\\
  & \le \frac{\delta^2}{8} \abspipes{
   \vtil_1^\top AH(x')A^\top \vtil_1 - \vtil_0^\top AH(x')A^\top\vtil_0
   }\\
  &~~~~+
   \frac{\delta^2}{8} \abspipes{
   \vtil_1^\top A[H(x_1')-H(x')]A^\top \vtil_1 - \vtil_0^\top A[H(x_0')-H(x')]A^\top\vtil_0
   }\\
  & = \frac{\delta^2}{8} \abspipes{
   v_1^\top[H(x_1')-H(x')]v_1 - v_0^\top [H(x_0')-H(x')]v_0
   }\\
  & \le \frac{\gamma\delta^3}{8}\left\{\norm{v_1}^3+\norm{v_0}^3\right\}.
 \end{align*}
 Here, the third line follows from the triangle inequality, the
 penultimate line from the fact that $A H(x') A=I_d$ and
 $\norm{\vtil_1}=\norm{\vtil_2}$, and the final line since
 $v^\top [H(b)-H(a)]v\le \gamma \norm{b-a} \norm{v}^2$.
\end{proof}

\section{Empirical exploration of the efficiency of \emph{Hug}}
\subsection{Optimal acceptance rate}
\label{sec.empirical.hug.eff}
We explore the relationship between the efficiency of Hug and the
acceptance rate by taking a grid of values for $T = 0.5, 1 \to 5$ and
$B = 1 , \ldots, 20$ on some example models in dimensions 25, 50, 75 and
100.  For each value of the tuple (Model, dimension, $B$, $T$), the
following procedure was performed for $i=1 \to 10,000$:

\begin{enumerate}
\item draw a value for $x_i$ directly from the target;
\item apply Hug with parameters $B$, $T$ to obtain $x_i'$;
\item record $N_i = \norm{x_i' - x_i}$ and $\alpha_i = \alpha(x_i', x_i)$.
\end{enumerate}

\begin{figure}[!ht]
  \centering
  \includegraphics[width=0.8\textwidth]{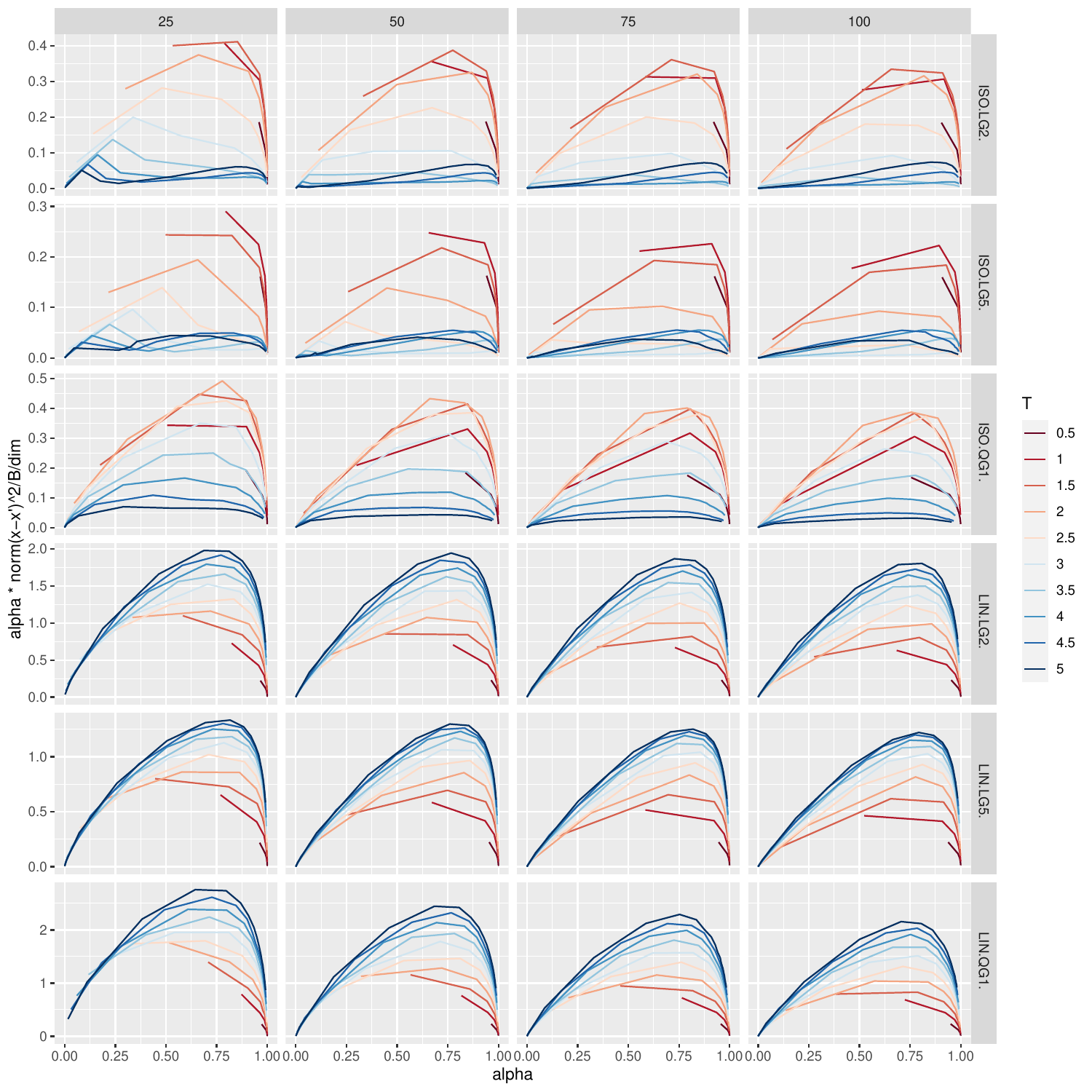}
 \caption{Efficiency plots for Hug: $\alpha$ vs
  $\EE{\alpha\norm{X'-X}^2}/dB$ for a range of $B$s (each line) on some
  example models (rows) in increasing dimensions(columns). For
  Isotropic targets $\sigma = 1$ and for LIN targets $\sigma_i =
  1 \to 10$. The forms for LG and QG targets can be found in
  Equation~(\ref{eqn.lga}) and (\ref{eqn.qga}) resp.
}
\label{fig:hug_opt_alpha}
\end{figure}

Figure~\ref{fig:hug_opt_alpha} shows the efficiency of Hug by plotting
$\hat{\mathbb{E}} [A N^2]/(dB)$ against acceptance rate $\alpha$; the y-axis
approximates the efficiency per unit time since the computational
effort for an iteration is essentially proportional to $B$; scaling by
$d$ is to compensate for the fact that when $x$ has $d$ components,
$\norm{x}^2 \propto d$.

%\clearpage

\subsection{Stability of hug}\label{sec:stability-hug}
We first describe a scenario which can be problematical for Hug, then explore a range of more typical scenarios. 

\begin{figure}[ht]
  \centering
 \includegraphics[width=0.7\textwidth]{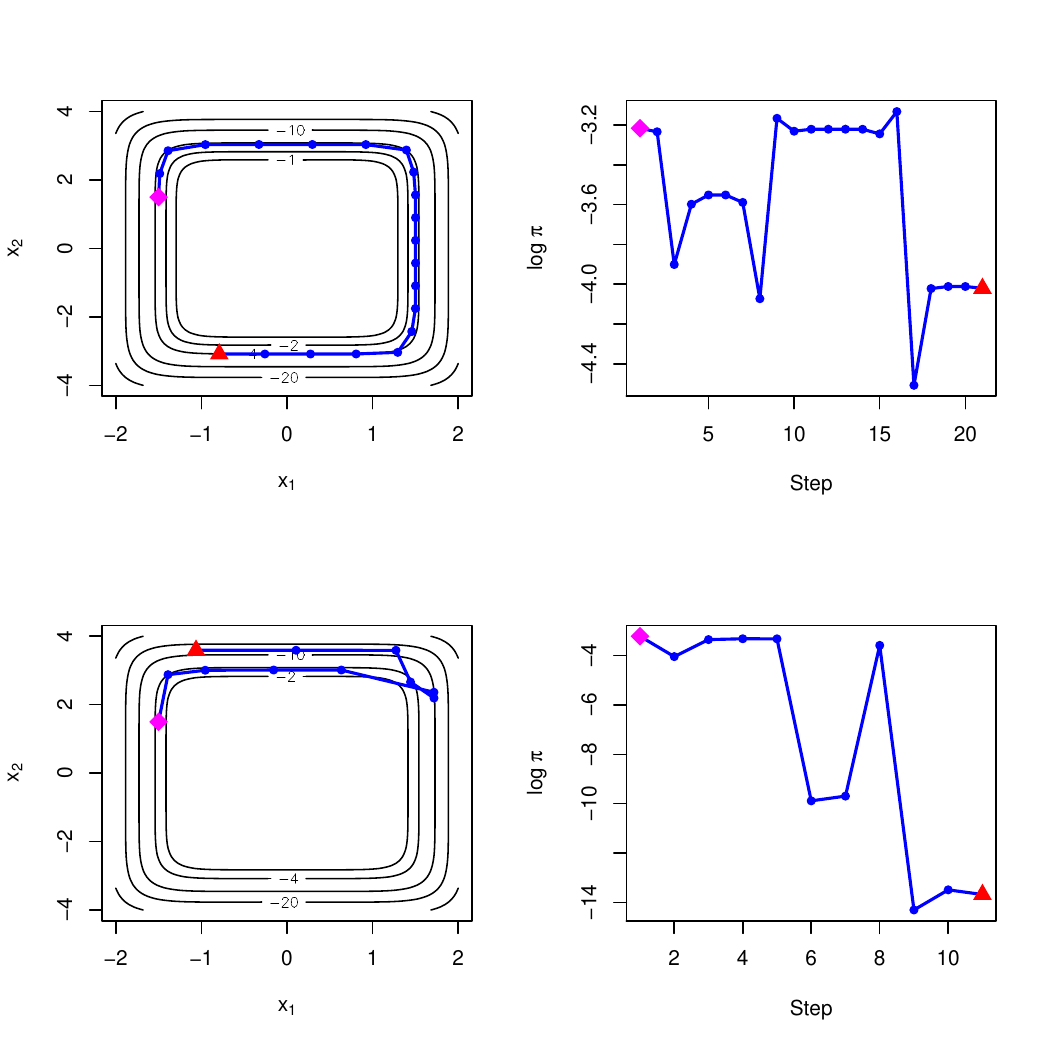}
 \caption{Left: two paths from the Hug algorithm on the target \eqref{eq.HugOnePath}, started at $(-1.5,1.5)$ (magenta diamond) with the same initial velocity drawn from $\Normal(0,I_2)$ and ending after $T=10$ time units (red triangle). Right: the corresponding values of $\log \pi$ at each point on the path. The top row corresponds to $B=20$ and the bottom row to $B=10$.}
\label{fig:HugOnA}
\end{figure}

In the following two-dimensional target the norms of the gradient and Hessian increase without bound as $||x||\rightarrow \infty$, contravening Conditions 1 and 2 of Theorem \ref{thm:hug-total-error}:
\begin{equation}
  \label{eq.HugOnePath}
\log \pi(x)=-\frac{1}{8}\left\{x_1^8+\left(\frac{x_2}{2}\right)^8\right\}+\mbox{constant}.
\end{equation}
 The relatively sharp ``corners'', where the curvature suddenly increases cause problems for Hug. The top row of Figure \ref{fig:HugOnA} shows that when $\delta$ is sufficiently small, the behaviour at any given contour can be controlled. This value of $\delta$ is much smaller than is necessary for good behaviour on the ``sides'' of the contours, which suggests increasing it; however, doubling $\delta$ leads to an unexpected path and a proposal that is very unlikely to be accepted.

To investigate this further, we created a $d=25$-dimensional target with:
\[
\log \pi_8(x)=-\frac{1}{8}\sum_{i=1}^d \frac{x_i^8}{\sigma_i^8}+\mbox{constant},
\]
and the scales $\sigma_i=1+2(i-1)/(d-1)$ and ran Hug for $T=4.0$ with $\delta$ ranging from $0.04$ to $0.8$. We started each run from a random point in the main posterior mass and noted the acceptance rate for Hug each time. The Hop parameters, $(\lambda,\kappa)$ were set to sensible values of $(0.5,1)$, which led to an acceptance rate for Hop of $\approx 40\%$, but no effort was made to tune them.

\begin{figure}[ht]
  \centering
 \includegraphics[width=0.5\textwidth]{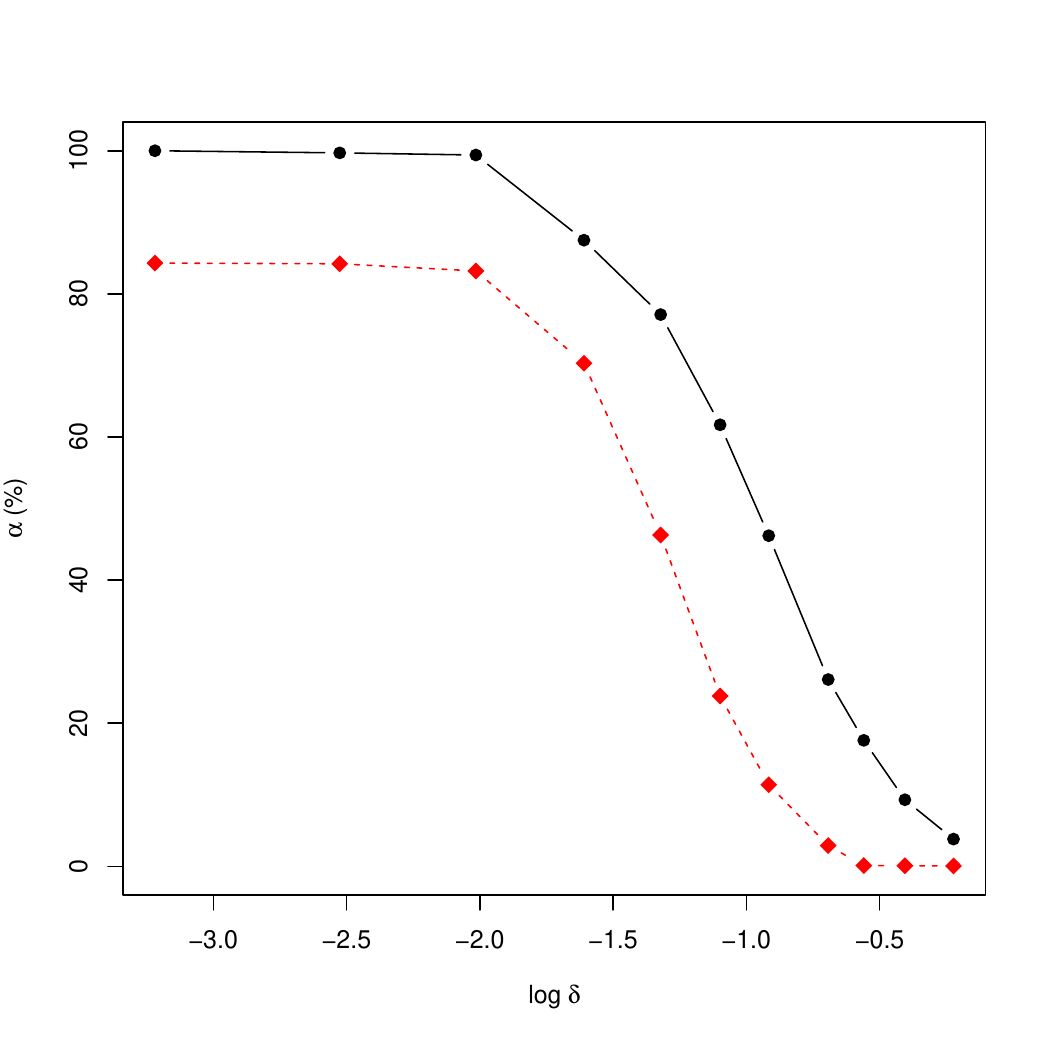}
 \caption{The empirical acceptance rate of Hug (solid, black curve) for various values of $\delta$, with $(T,\lambda,\kappa)=(4,1,0.5)$, and Hamiltonian Monte Carlo (dotted, red curve) for the same $\epsilon\equiv \delta$ values with $T=4$.}
\label{fig:HugOnAacc}
\end{figure}

The black curve in Figure \ref{fig:HugOnAacc} shows how the acceptance rate for Hug plummets as $\delta$ is increased. For comparison, we also ran Hug and Hop on the Targets $\pi_G$ (Gaussian) and $\pi_{LG}$ and $\pi_{QG}$ of Section \ref{sec:sims} but with the same set of $\sigma_i$ as here, and using the same tuning parameters as here but with the largest $\delta$, $0.8$. Even though the target scales are similar, because of the lack of sharp corners, especially with the first two targets, the acceptance rates were much higher: respectively, 94.4\%, 97.5\% and 63.0\%. The quartic terms in the third target caused some deterioriation.

Hamiltonian Monte Carlo suffers even more drastically in this situtation. With $(T,\delta)=(4.0,0.8)$, the acceptance rates for the Targets $\pi_G$ (Gaussian) and $\pi_{LG}$ and $\pi_{QG}$ were respectively, 91.1\%, 95.2\% and 37.8\%, indicating that this is again a reasonable scaling, but that performance is more substantially reduced when the target has quartic terms and the hints of corners start to appear. The dotted, red curve in Figure \ref{fig:HugOnAacc} shows the acceptance rate for Hamiltonian Monte Carlo applied to $\pi_8$ using the set of $\delta$ values also used for Hug. Not only does the acceptance rate reduce to effectively zero much earlier, but no matter how small $\delta$ is, the acceptance rate cannot be increased above about 84\%. Much of the posterior for the $i$th component is contained within
$[-1.5\sigma_i,1.5 \sigma_i]$, but at the edges of this range the magnitude of the gradient is $1.5^7/\sigma_i\approx 17/\sigma_i$. For the lower $\sigma_i$ the Leapfrog scheme will only produce a sensible path if the velocity component in these directions is small. As discussed in Section \ref{sec.ergANDconv}, Hug only depends on the direction of the gradient, not its magnitude and so, is relatively stable compared with this behaviour.

\begin{figure}[!ht]
  \centering
 \includegraphics[width=0.8\textwidth]{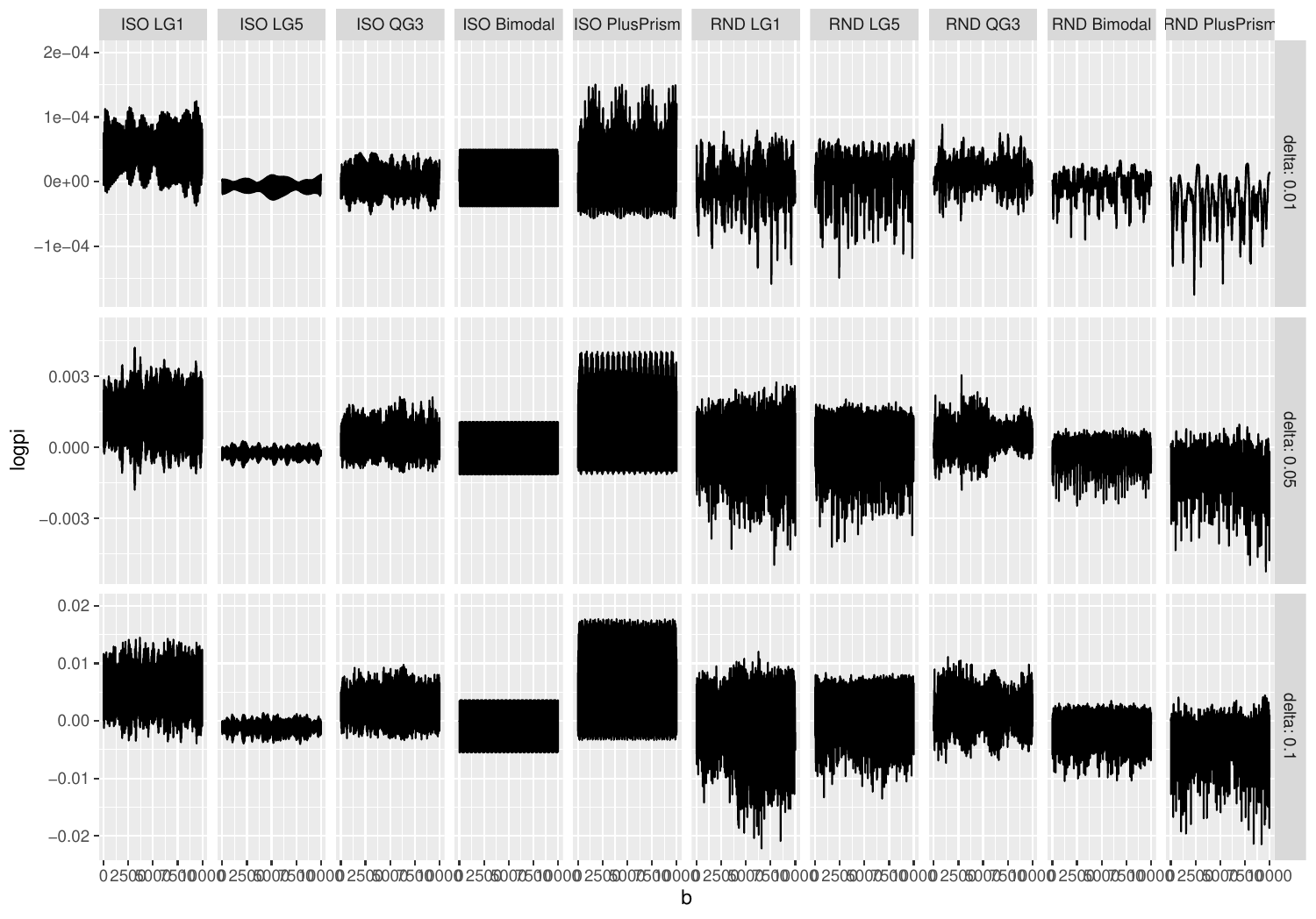}
 \caption{ Stability of $\ell$ for various 25-dimensional models
   (columns) under the \emph{Hug} algorithm with $\delta = 0.01, 0.05$
   and $0.1$ (rows). Each plot shows
   $\Delta_i = \ell(x_i) - \ell(x_0)$ vs.  $i = 0 \to 10,000$. }
\label{fig:hug-stability}
\end{figure}

Figure \ref{fig:hug-stability} shows a plot of
$\ell(x_b) - \ell(x_0)$ against iteration number $b$ of the inner loop
in Hug algorithm for a range of 25-dimensional
models (definitions for which can be found in the main article,
Section~\ref{sec:sims}). \emph{Iso} models have all scales equal to 1
while \emph{Rand} models have scales simulated from U(1, 5). The
limits on the $y$-axis are chosen as double the maximum and minimum of
$\ell(x_b) - \ell(x_0)$.

\section{Additional material for Hop}
\label{app.add.Hop}
The inverse and square-root of $B_x$ are as follows:
\begin{align}
 \label{eqn.Binv}
 B_x^{-1} & = \frac{1}{\mu^2} I + \left(\frac{1}{\lambda^2} -
 \frac{1}{\mu^2}\right) \ghat(x)\ghat(x)^\top, \\
 \label{eqn.Bsqrt}
 B_x^{1/2} & = \mu I + (\lambda-\mu) \ghat(x)\ghat(x)^\top.
\end{align}

For the standard version of Hop, the acceptance ratio in \eqref{eq:hop-log-accept} simplifies to:
\begin{align}
 \nonumber
 \log r_{hop}(x,y) & = \ell(y) - \ell(x) + \log \Normal\left(x; y, B_y/\norm{g(y)}^2\right) - \log \Normal\left(y; x, B_x/\norm{g(x)}^2\right) \\
 & = \ell(y) - \ell(x) + \frac{d}{2} \log\frac{\norm{g(y)}^2}{\norm{g(x)}^2} - \frac{1}{2}\left(y-x\right)^\top\left[ \norm{g(y)}^2 B_y^{-1} - \norm{g(x)}^2 B_x^{-1} \right]\left(y-x\right),\nonumber \\
 & = \ell(y) - \ell(x) + \frac{d}{2} \log\frac{\norm{g(y)}^2}{\norm{g(x)}^2} - \frac{1}{2\mu^2} \norm{y-x}^2\left[\norm{g(y)}^2-\norm{g(x)}^2\right]\nonumber \\
 & \qquad- \frac{1}{2}\left(\frac{1}{\lambda^2}-\frac{1}{\mu^2}\right)\left\{[\left(y-x\right)^\top\!g(y)]^2 -[\left(y-x\right)^\top\!g(x)]^2\right\}.\label{eq:hop-log-accept-messy}
\end{align}
using \eqref{eqn.Binv}, and since $\mathsf{det}(B_x)=\lambda\mu^{d-1}$
is independent of $x$.

For position-specific preconditioning of Hop,
$\norm{\tilde g(x)}^2 = g(x)^\top \Sigma(x) g(x)$ and the algorithm proposes points:
\begin{align}
 \label{eq:hop-hessian}
 Y|X = x \sim \Normal\left(x, \frac{1}{g(x)^\top \Sigma(x) g(x)} \left(\mu^2 \Sigma(x) + (\lambda^2 - \mu^2) \frac{\Sigma(x) g(x) g(x)^\top \Sigma(x)^\top}{g(x)^\top \Sigma(x) g(x)} \right)\right)
\end{align}

For a proposed point $y$ from $x$, the log-acceptance ratio for a Hop
using Hessian information is:
\begin{align}
 \label{eq:hop-hess-log-accept}
 \log r_{hopH} & = \ell(y) - \ell(x) + \frac{d}{2} \log \frac{\norm{\tilde g(y)}^2}{\norm{\tilde g(x)}^2} + \frac{1}{2}\log\frac{\det{\Sigma(x)}}{\det{\Sigma(y)}} \nonumber \\
 & \quad - \frac{1}{2\mu^2}\left(y-x\right)^\top \left( \norm{\tilde g(x)}^2H_x - \norm{\tilde g(y)}^2H_y\right)\left(y-x\right) \\
 & \quad - \frac{1}{2}\left(\frac{1}{\lambda^2} -\frac{1}{\mu^2}\right)\left( [\left(y-x\right)^\top g(y)]^2 - [\left(y-x\right)^\top g(x)]^2\right)\nonumber
\end{align}

\section{Proof of Theorem~\ref{thm:hop_prod}}
\label{app:hop_prod_proof}

\subsection{Notation and definitions}
In proving Theorem~\ref{thm:hop_prod} we drop the superscript $^{(d)}$ from $\pi^{(d)}$ and $x^{(d)}$ and the subscript from $\lambda_d$ and $\mu_d$ whenever this is clear from the context. Let $\ell=\log \pi$, $g=\nabla \ell$, $g_1=\ell_1'$ and for $X$ with a density of $\exp(\ell_1)$, define
$$
\begin{aligned}
m_2&=\Expect{g_1(X)^2}=\Expect{-g_1'(X)}.
\end{aligned}
$$
We switch between the following equivalent forms for the proposed jump vector: 
$$
y-x=\frac{1}{||g(x)||}(\lambda \zpara+\mu\zperp)=\frac{1}{||g(x)||}\{(\lambda-\mu)\zpara+\mu z\},
$$

where $Z\sim \Normal(0,I_d)$ and $\Zpara=\{\ghat(x)^\top Z\}\ghat(x)$ is a $\Normal(0,1)$ variable along the $\ghat(x)$ vector and $\zperp=z-\zpara$. Clearly $||\zperp||=\mathcal{O}(d^{1/2})$, whereas $||\zpara||=\mathcal{O}(1)$. 

Further, for $U\sim \Normal(0,1)$ we define $e_a:=\Expect{|U|^a}$ and $v_a:=\Var{|U|^a}$.

Throughout, "$\to$" indicates convergence in probability when associated with a sequence of random variables. Also, for a sequence of random variables $A_d$ and a sequence of real numbers $b_d$, we write $A_d\sim b_d$ iff $A_d/b_d\to 1$ and $A_d\lesssim b_d$ iff $A_d/b_d\to c$ for some $c\in [0,1]$.

It is natural to split the log acceptance ratio \eqref{eq:hop-log-accept-messy} into four terms:

$$
\begin{aligned}
C_1&=\ell(y)-\ell(x)\\
C_2&=\frac{d}{2}\log \frac{||g(y)||^2}{||g(x)||^2},\\
C_3&=-\frac{1}{2\mu^2}||y-x||^2\left\{||g(y)||^2-||g(x)||^2\right\},\\
C_4&=-\frac{1}{2}\left(\frac{1}{\lambda^2}-\frac{1}{\mu^2}\right)
\left[\left\{(y-x)^\top g(y)\right\}^2-\left\{(y-x)^\top g(x)\right\}^2\right]
=
C_4^a C_4^b,
\end{aligned}
$$
where
$$
C_4^a=(y-x)^\top ~\frac{g(y)+g(x)}{2}.
$$
and
$$
C_4^b=-\left(\frac{1}{\lambda^2}-\frac{1}{\mu^2}\right)
(y-x)^\top \{g(y)-g(x)\}.
$$

The typical sizes of all four terms increase without bound as $d$ increases. However, note that
$$
C_2
=
\frac{d}{2}\log \left\{1+\frac{||g(y)||^2-||g(x)||^2}{||g(x)||^2}\right\}
\approx
\frac{d}{2}\frac{||g(y)||^2-||g(x)||^2}{||g(x)||^2}
$$
and
$$
C_4^a\approx \ell(y)-\ell(x).
$$
We will show that $C_2$ and $C_3$ cancel except for terms vanishingly small in $d$, and that the only non-vanishing remainder from $C_1+C_4$ is $\mathcal{O}(1)$ and leads to the stated acceptance ratio.

\subsection{Elementary results}

We first gather together some elementary results that will be used repeatedly.

\begin{proposition}
\label{prop.Zpara.mom}
  $$
\begin{aligned}
\Expect{\Zpara_i Z_i}
=
%\Expect{\sum_{j=1}^d\frac{g_1(x_j)}{||g(x)||} Z_j \frac{g_1(x_i)}{||g(x)||}~Z_i}
%=
\frac{g_1(x_i)^2}{||g(x)||^2}
=\Expect{(\Zpara_i)^2},\\
\Expect{\Zpara_i Z_i^2}
=
\Expect{(\Zpara_i)^2 Z_i}
=\Expect{(\Zpara_i)^3}=0,\\
\Var{\sum_{i=1}^d\{(\lambda-\mu)\Zpara_i +\mu Z_i\}^a h(x_i)}
\sim
k_a d\mu^{2a}\Expect{h(X)^2},
\end{aligned}
$$
for $a=2,3$, where $k_2=2$ and $k_3=15$.
\end{proposition}
\begin{proof}
We prove the final result for $a=3$; a similar method gives the result with $a=2$. Firstly
  \[
  d^{1/2}\sum_{i=1}^d \ghat_i(x)^3h(x_i)=\frac{d^{1/2}}{||g(x)||^3}\sum_{i=1}^d g_1(x_i)^3 h(x_i)
  \rightarrow \frac{\Expect{g_1(X)^3h(X)}}{m_2^{3/2}},
  \]
  so $\sum_{i=1}^d \ghat_i^3h(x_i)=\mathcal{O}(1/d^{1/2})$. Analogously, $\sum_{i=1}^d \ghat_i^4h(x_i)=\mathcal{O}(1/d)$ and $\sum_{i=1}^d \ghat_i^2h(x_i)=\mathcal{O}(1)$, with similar outcomes if $h$ is replaced with $h^2$.  Now
  \[
  \sum_{i=1}^d \{(\lambda-\mu) \Zpara_i+\mu Z_i\}^3h(x_i)
  =A_1(x,Z)+A_2(x,Z)+A_3(x,Z)+A_4(x,Z).
  \]
  Here, because $\Zpara=(Z\cdot \ghat)\ghat=Z_*\ghat$, where $Z_*\sim \Normal(0,1)$ marginally,
  \begin{align*}
    A_1(x,Z)
    &=(\lambda-\mu)^3Z_*^3\sum_{i=1}^d\ghat_i^3h(x_i)
  =
  \mathcal{O}(\lambda^3/d^{1/2}),\\
  A_2(x,Z)&=3\mu(\lambda-\mu)^2Z_*^2 \sum_{i=1}^d \ghat_i^2 h(x_i)Z_i 
  =3\mu(\lambda-\mu)^2 Z_*^2 ~\Normal\left(0,\sum_{i=1}^d \ghat_i^4h^2(x_i)\right),\\
  A_3(x,Z)&=3\mu^2(\lambda-\mu) Z_*\sum_{i=1}^d \ghat_ih(x_i) Z_i^2,\\
  A_4(x,Z)&=\mu^3\sum_{i=1}^d Z_i^3h(x_i).
  \end{align*}
  For any two random variables, $A$ and $B$, the Cauchy-Schwarz inequality gives $\Var{AB}\le \Expect{A^2B^2}\le \Expect{A^4}^{1/2}\Expect{B^4}^{1/2}$.
  So \begin{align*}
    \Var{A_1}&=\mathcal{O}(\lambda^6/d),\\
  \Var{A_2}&\le 9\mu^2\lambda^4\times \mathcal{O}(1/d^2)=\mathcal{O}(\lambda^5/d),\\
\Var{A_3}&\le 9\mu^4\lambda^2\times \mathcal{O}(1)=\mathcal{O}(\lambda^4),\\
\Var{A_4}&=15\mu^6d=\mathcal{O}(\lambda^3 d).
  \end{align*}
Now $\Var{A_1}/\Var{A_4}=\mathcal{O}(1/d^{2})\rightarrow 0$, $\Var{A_2}/\Var{A_4}=\mathcal{O}(\lambda^2/d^2)\rightarrow 0$, and $\Var{A_3}/\Var{A_4}=\mathcal{O}(\lambda/d)\rightarrow 0$. So even if the correlations between $A_1$, $A_2$, $A_3$ and $A_4$ were all $1$, the only important variance term asymptotically would be that of $A_4$. The result follows as the $Z_i$ are independent and $\Var{Z_i^3}=15$.
  \end{proof}

\begin{proposition}
  \label{prop.gstuff}
  $$
  \begin{aligned}
    \frac{1}{d}||g(X)||^2
    &=\frac{1}{d}\sum_{i=1}^d g_1(X_i)^2\to m_2,\\
||g(y)-g(x)||
&
%=
%\left[\sum_{i=1}^d\{g_1(y_i)-g_1(x_i)\}^2\right]^{1/2}
%\le
%\left\{\sum_{i=1}^dT^2(y_i-x_i)^2\right\}^{1/2}
%=
\le
T||y-x||.
  \end{aligned}
  $$
  \end{proposition}

\subsection{$C_2$ and $C_3$ cancel}

\begin{lemma}
  \label{lem.prelim}
  \[
g(x)^\top\{g(y)-g(x)\}\lesssim e_1 T \lambda_d. 
  \]
\end{lemma}
\begin{proof}
By Taylor expansion,
$g(x)^\top\{g(y)-g(x)\}=W_1+W_2$, where 
$$
W_1(x,z)=\frac{\lambda-\mu}{||g(x)||}
\sum_{i=1}^d g_1(x_i)g_1'(x^*_i)~ \zpara_i
~~~
\mbox{and}
~~~
W_2(x,z)=\frac{\mu}{||g(x)||}
\sum_{i=1}^d g_1(x_i)g_1'(x^*_i)~z_i.
$$
for some $x^*(x,z)=t x + (1-t) y$, $0\le t(x,z) \le 1$. Now for large $d$, $\lambda>\mu$ and
$$
\begin{aligned}
|W_1(x,z)|&\le B_1(x,z):=\frac{\lambda T}{||g(x)||} \sum_{i=1}^d |g_1(x_i)|~|\zpara_i|,\\
|W_2(x,z)|&\le B_2(x,z):=\frac{\mu T}{||g(x)||}\sum_{i=1}^d|g_1(x_i)|~|z_i|.
\end{aligned}
$$
Since $\Zpara_i\sim \Normal(0,g_1(x_i)^2/||g(x)||^2)$ are independent, 
$$
\begin{aligned}
\Expect{B_1(x,Z)}
&=
\frac{\lambda Te_1}{||g(x)||^2}
\sum_{i=1}^d g_1(x_i)^2
= T e_1 \lambda ,\\
d \Var{B_1(x,Z)}
&=
d \frac{\lambda^2 T^2}{||g(x)||^4}v_1\sum_{i=1}^d g_1(x_i)^4
\to
\frac{T^2 m_4 v_1\lambda^2 }{m_2^2}.
\end{aligned}
$$
So $|W_1|/\lambda_d \le B_1/\lambda_d \to T e_1$. Similarly,
$$
\begin{aligned}
\Expect{B_2(x,Z)^2}
&\le
\frac{\mu^2T^2}{||g(x)||^2}
\sum_{i=1}^d
g_1(x_i)^2\Expect{Z_i^2}
=
\mu^2T^2,\\
\Expect{B_2(x,Z)^4}
&\le
\frac{\mu^4T^4}{||g(x)||^4}
\sum_{i=1}^d
g_1(x_i)^4\Expect{Z_i^4}
+
\frac{\mu^4T^4}{||g(x)||^4}
\sum_{i=1}^d
\sum_{j=1}^d
g_1(x_i)^2g_1(x_j)^2\Expect{Z_i^2}\Expect{Z_j^2}\\
&=
\frac{2\mu^4T^4}{||g(x)||^4}\sum_{i=1}^d
g_1(x_i)^4
+
\Expect{B_2^2}^2
\end{aligned}
$$
The variance term is:
$$
\frac{2\mu_d^4T^4}{||g(x)||^4}\sum_{i=1}^d
g_1(x_i)^4
\sim
\frac{2\kappa^2 T^4 m_4\lambda_d^2}{m_2^2d}\rightarrow 0,
$$
by \eqref{eqn.lambda.behave}.
So $B_2^2/\mu_d^2 \to T^2$, and $|W_2|/\mu_d\le B_2/\mu_d\to T$. The result follows since $\mu_d/\lambda_d\rightarrow 0$.
  \end{proof}

\begin{lemma}
  \label{lemma.vanishing}
  \[
  d^{1/2}\times \frac{||g(y)||^2-||g(x)||^2}{||g(x)||^2}\to 0.
  \]
\end{lemma}
\begin{proof}
Firstly,
$$
||g(y)||^2-||g(x)||^2
=
||g(y)-g(x)||^2+2g(x)^\top\{g(y)-g(x)\}.
$$
Since $||\Zpara||/d^{1/2}\downarrow 0$ and $||\Zperp||/d^{1/2}\rightarrow 1$, Proposition \ref{prop.gstuff} and \eqref{eqn.lambda.behave} give:
$$
||g(y)-g(x)||^2\le T^2\frac{\lambda^2|| \Zpara||^2+\mu^2 ||\Zperp||^2}{||g(x)||^2}
\sim T^2\frac{\mu^2}{m_2}=\frac{T^2\kappa\lambda}{m_2},
$$
Secondly, by Lemma \ref{lem.prelim} $g(x)^\top\{g(y)-g(x)\}\sim e_1T\lambda$.
So, combining,
$$
d^{1/2}\times \frac{||g(y)||^2-||g(x)||^2}{||g(x)||^2}
\lesssim
d^{-1/2}\frac{T^2\kappa +2e_1T}{m_2}\lambda_d\to 0
$$
by \eqref{eqn.lambda.behave}.
\end{proof}

Lemma \ref{lemma.vanishing} will be used several times, the first of which is in the Taylor expansion of $C_2$.

\begin{corollary}
$$
\frac{d}{2}\log \frac{||g(y)||^2}{||g(x)||^2}
-
\frac{d}{2} \frac{||g(y)||^2-||g(x)||^2}{||g(x)||^2}\rightarrow 0 
$$
as $d\rightarrow \infty$.
\end{corollary}
\begin{proof}
  Firstly
  \[
C_2=\frac{d}{2}\log\left\{1+\frac{||g(y)||^2-||g(x)||^2}{||g(x)||^2}\right\}. 
\]
By Lemma \ref{lemma.vanishing} the Taylor expansion of $C_2$ can be made absolutely convergent with a probability as close to $1$ as desired by taking a large enough $d$. Finally, the $a$th term in the expansion is
\[
\frac{d}{2}\times \left(\frac{||g(y)||^2-||g(x)||^2}{||g(x)||^2}\right)^a
=\frac{1}{2}\times \left(d^{1/a}\times \frac{||g(y)||^2-||g(x)||^2}{||g(x)||^2}\right)^a\to 0
\]
for $a\ge 2$ by Lemma \ref{lemma.vanishing}.
\end{proof}

\begin{corollary}
  \[
C_2+C_3\to 0~~~\mbox{as}~~~d\rightarrow \infty.
\]
\end{corollary}
\begin{proof}
  Define
$$
C_3^a=-\frac{1}{2\mu^2}\frac{\lambda^2 ||\Zpara||^2}{||g(x)||^2}
\left\{||g(y)||^2-||g(x)||^2\right\}
=
-\frac{\lambda}{2\kappa} ||\Zpara||^2\times \frac{||g(y)||^2-||g(x)||^2}{||g(x)||^2}\rightarrow 0,
$$
as $d\rightarrow \infty$ by Lemma \ref{lemma.vanishing} because $\lambda=o(d^{1/2})$ and $||\Zpara||=\mathcal{O}(1)$. Also define
$$
C_3^b=-\frac{1}{2\mu^2}\frac{\mu^2 ||\Zperp||^2}{||g(x)||^2}\left\{||g(y)||^2-||g(x)||^2\right\}
=
-\frac{1}{2} ||\Zperp||^2\times \frac{||g(y)||^2-||g(x)||^2}{||g(x)||^2}.
$$
So, ignoring those terms in $C_2$ that we already know vanish,
$$
C_2+C_3^b=\frac{1}{2}\frac{||g(y)||^2-||g(x)||^2}{||g(x)||^2}\left\{d-||\Zperp||^2\right\}\rightarrow 0
$$
as $d\rightarrow \infty$ by Lemma \ref{lemma.vanishing} because $d-||\Zperp||^2=\mathcal{O}(d^{1/2})$. The result follows since $C_3=C_3^a+C_3^b$.
\end{proof}

\subsection{Expanding $C_1$ to give the bottom of \eqref{eqn.acclargeDelta}}

\begin{lemma}
  \label{lemma.provelargeDelta}
The lower half of \eqref{eqn.acclargeDelta} holds; \emph{i.e.},
  \[
\frac{1}{\lambda_d} \left\{\log \pi^{(d)}\left(Y^{(d)}\right)-\log\pi^{(d)}\left(X^{(d)}\right)\right\}\Longrightarrow
\Normal\left(-\frac{1}{2}\kappa,1\right).
  \]
  \end{lemma}

\begin{proof}
A third-order Taylor expansion gives:
$$
\begin{aligned}
\frac{1}{\lambda}\{\ell(y)-\ell(x)\}
&=
\frac{1}{\lambda}(y-x)^\top g(x)
+
\frac{1}{2\lambda}\sum_{i=1}^d (y_i-x_i)^2g_1'(x_i)
+
\frac{1}{6\lambda}\sum_{i=1}^d (y_i-x_i)^3g_1''(x_i)\\
&~~~+\frac{1}{6\lambda}\sum_{i=1}^d (Y_i-x_i)^3\{g_1''(x_i^*)-g_1''(x_i)\}\\
&=A_1(x,z)+A_2(x,z)+A_3(x,z)+A_4(x,z),
\end{aligned}
$$
for some $x^*=x+ty$ and some $t(x,y)\in [0,1]$.
Here $A_1(x,Z)=(\Zpara)^\top \ghat\sim \Normal(0,1)$ and
$$
\begin{aligned}
A_4(x,z)
&=
\frac{1}{6\lambda ||g(x)||^3}\sum_{i=1}^d \left\{(\lambda-\mu)\zpara_i+\mu z_i\right\}^3\{g_1''(x_i^*)-g_1''(x_i)\}\\
&\le
\frac{L}{6\lambda ||g(x)||^4}\sum_{i=1}^d\left|(\lambda-\mu)\zpara_i+\mu z_i\right|^4 \sim \frac{L}{6\lambda d^2 m_2^2}\times 3d \mu^4
=
\frac{L\kappa ~\lambda_d}{2 m_2^2 ~d}\rightarrow 0,
\end{aligned}
$$
by \eqref{eqn.lambda.behave}. Now, $A_3(x,z)=\frac{1}{6\lambda ||g(x)||^3}\sum_{i=1}^d \left\{(\lambda-\mu)\zpara_i+\mu z_i\right\}^3g_1''(x_i)$, so $\Expect{A_3(x,Z)}=0$ by Proposition \ref{prop.Zpara.mom}. By the same proposition, 
$$
\Var{A_3(x,Z)}\sim 
\frac{15 \mu^6 d}{36\lambda^2 d^3m_2^3}\times  \Expect{g_1''(X)^2}
=
\frac{5\kappa^3~\lambda_d}{12m_2^3~d}\Expect{g_1''(X)^2}\rightarrow 0,
$$
so the third term also vanishes. Finally, the second term is
$$
A_2(x,z)=\frac{1}{2\lambda ||g(x)||^2}\sum_{i=1}^d \left\{(\lambda-\mu)\zpara_i+\mu z_i\right\}^2g_1'(x_i).
$$
By Proposition \ref{prop.Zpara.mom}, it satisfies:
$$
\begin{aligned}
\Expect{A_2(x,Z)}
&=
\frac{\kappa}{2||g(x)||^2}
\sum_{i=1}^d \Expect{\left\{\left(\frac{\lambda}{\mu}-1\right)\Zpara_i+Z_i\right\}^2} g_1'(x_i)\\
&=
\frac{\kappa}{2||g(x)||^2}
\sum_{i=1}^d \left(\frac{\lambda^2}{\mu^2}-1\right)\frac{g_1(x_i)^2}{||g(x)||^2}g_1'(x_i)+g_1'(x_i)
\sim
\frac{\kappa}{2 m_2}\Expect{g_1'(X)}=-\frac{\kappa}{2},
\end{aligned}
$$
since $\lambda^2/\mu^2=\lambda/\kappa=o(d^{1/2})$. Finally, by Proposition \ref{prop.Zpara.mom},
$$
\Var{A_2(x,\mu)}
\sim
\frac{2\mu^4 d}{4\lambda^2m_2^2d^2}\Expect{g_1'(X)^2}=
\frac{\kappa^2}{2 d m_2^2}\Expect{g_1'(X)^2}\rightarrow 0.
$$
\end{proof}

\subsection{The terms $C_1$ and $C_4$}

\begin{lemma}
  \label{lem.trapeze}
$C_4^a-\{\ell(y)-\ell(x)\}\to 0$ as $d\rightarrow \infty$.
\end{lemma}
\begin{proof}
By a similar error analysis as used in the trapezoidal rule,
$$
C_4^a-\{\ell(y)-\ell(x)\}
=\frac{1}{12}\sum_{i=1}^d (y_i-x_i)^3g_1''(x_i^*)
=D_1+D_2,
$$
where
$$
\begin{aligned}
D_1&=\frac{1}{12}\sum_{i=1}^d (y_i-x_i)^3g_1''(x_i)
=\frac{1}{12||g(x)||^3}\sum_{i=1}^d \left\{(\lambda-\mu)\zpara_i+\mu z_i\right\}^3g_1''(x_i),\\
D_2&=\frac{1}{12}\sum_{i=1}^d (y_i-x_i)^3\{g_1''(x_i^*)-g_1''(x_i)\}.
\end{aligned}
$$
Now $\Expect{D_1(x,Z)}=0$ by Proposition \ref{prop.Zpara.mom}, and  by the same proposition, 
$$
\begin{aligned}
\Var{D_1(x,Z)}\sim \frac{15\mu^6 d}{144 m_2^3 d^3} \Expect{g_1''(X)^2}
=
\frac{5\kappa^3~\lambda_d^3}{48 m_2^3~d^2}\times \Expect{g_1''(X)^2}
\rightarrow 0
\end{aligned}
$$
as $d\rightarrow \infty$ because $\lambda=o(d^{1/2})$; thus $D_1\to 0$. Also,
$$
\begin{aligned}
|D_2(x,z)|&\le \frac{L}{12}\sum_{i=1}^d |y_i-x_i|^4
=
\frac{L}{12||g(x)||^4}\sum_{i=1}^d |\lambda \zpara_i+\mu \zperp_i|^4\\
&\sim
\frac{L}{12d^2 m_2^2}\times 3d \mu^4
=
\frac{L\kappa^2 ~\lambda_d^2}{4 m_2^2~d}\rightarrow 0
\end{aligned}
$$
as $\lambda_d^2/d\rightarrow 0$.
\end{proof}

Since $C_4^a\sim \ell(y)-\ell(x)=\mathcal{O}(\lambda)$ we may neglect all terms in $C_4^b$ which are $o(1/\lambda)$.

\begin{lemma}
\label{lem.Cfourb} $\lambda_d\left\{C_4^b(X,Z)+1\right\}\to  \kappa$, with a discrepancy which is $o(1/\lambda_d^2)$.
  \end{lemma}

\begin{proof}
$$
C_4^b=\left(\frac{1}{\mu^2}-\frac{1}{\lambda^2}\right)\sum_{i=1}^d (y_i-x_i)\{g_1(y)-g_1(x)\}
=V_1+V_2+V_3
$$
where, by a second-order Taylor expansion,
$$
\begin{aligned}
V_1(x,z)&=\left(\frac{1}{\mu^2}-\frac{1}{\lambda^2}\right)\frac{1}{||g(x)||^2}\sum_{i=1}^d \{(\lambda-\mu) \zpara_i+\mu z_i\}^2g'_1(x_i),\\
V_2(x,z)&=\left(\frac{1}{\mu^2}-\frac{1}{\lambda^2}\right)\frac{1}{2||g(x)||^3}\sum_{i=1}^d \{(\lambda-\mu) \zpara_i+\mu z_i)^3g''_1(x_i),\\
V_3(x,z)&=\left(\frac{1}{\mu^2}-\frac{1}{\lambda^2}\right)\frac{1}{2||g(x)||^3}\sum_{i=1}^d \{(\lambda-\mu) \zpara_i+\mu z_i\}^3\{g''_1(x_i^*)-g''_1(x_i)\}.
\end{aligned}
$$
We first show that $V_2$ and $V_3$ are $o(1/\lambda_d)$. Proposition \ref{prop.Zpara.mom} gives $\Expect{V_2(x,Z)}=0$ and 
$$
\begin{aligned}
\Var{V_2(x,Z)}\sim \frac{1}{\mu^4}\frac{15\mu^6 d}{4m_2^3 d^3}\Expect{g_1''(X)^2}
=
\frac{15\kappa \lambda_d}{4m_2^3d^2}\Expect{g_1''(X)^2},
\end{aligned}
$$
So $V_2=\mathcal{O}(\lambda_d^{1/2}/d)=o(1/d^{3/4})$. For large $d$, $\lambda>\mu$, so
$$
|V_3|\le
\frac{L}{2\mu^2||g(x)||^4}\sum_{i=1}^d\{(\lambda-\mu) \zpara_i+\mu z_i\}^4
\sim
\frac{L \times 3\mu^4 d}{2\mu^2d^2m_2^2}
=
\frac{3L \kappa ~\lambda_d}{2m_2^2~d}=\mathcal{O}(\lambda_d/d)=o(1/\lambda_d).
$$

Finally, we tackle $V_1$. By Proposition \ref{prop.Zpara.mom} 
$$
\Var{V_1(x,Z)}\sim
\frac{2\mu^4 d}{\mu^4d^2m_2^2}\Expect{g_1'(X)^2}
=
\frac{2}{dm_2^2}\Expect{g_1'(X)^2},
$$
so variations from $\Expect{V_1(x,Z)}$ are $\mathcal{O}(1/d^{1/2})=o(1/\lambda_d)$ and can be neglected. Finally,
$$
\begin{aligned}
\Expect{V_1(x,Z)}
&=
\left(\frac{1}{\mu^2}-\frac{1}{\lambda^2}\right)\frac{1}{||g(x)||^2}
\sum_{i=1}^d
\left\{(\lambda^2-\mu^2)\frac{g_1(x_i)^2}{||g(x)||^2}+\mu^2\right\}g_1'(x_i).
\end{aligned}
$$
Now
$$
\left(\frac{1}{\mu^2}-\frac{1}{\lambda^2}\right)
\frac{1}{||g(x)||^2}
\sum_{i=1}^d
(\lambda^2-\mu^2)\frac{g_1(x_i)^2}{||g(x)||^2}g_1'(x_i)
<
\frac{\lambda^2}{\mu^2||g(x)||^4}
\sum_{i=1}^d
g_1(x_i)^2g_1'(x_i)
\sim
\frac{\lambda}{\kappa d m_2^2}\Expect{g_1(X)^2g_1'(X)},
$$
which is $\mathcal{O}(\lambda_d/d)$ and may be neglected. Hence we need only consider
$$
\left(\frac{1}{\mu^2}-\frac{1}{\lambda^2}\right)
\frac{1}{||g(x)||^2}
\sum_{i=1}^d
\mu^2 g_1'(x_i)
=
\left(1-\frac{\kappa}{\lambda}\right)
\frac{\sum_{i=1}^d g_1'(x_i)}{||g(x)||^2}
\sim
\frac{\kappa}{\lambda}-1,
$$
with a multiplicative error of $\mathcal{O}(1/d^{1/2})$, which can be neglected. So $C_4^b\sim \kappa/\lambda_d -1$ with errors of $o(1/\lambda_d)$, as required.
\end{proof}

\subsection{Proof of \eqref{eqn.limiting.acc.rate}}
Combining Lemmas \ref {lemma.provelargeDelta}, \ref{lem.trapeze} and \ref{lem.Cfourb}, after cancellation with $C_1$, the only terms arising from $C_4$ are
$$
\frac{\kappa}{\lambda}
\{\ell(y)-\ell(x)\}
\sim
\frac{\kappa}{\lambda}\lambda\left(U-\frac{\kappa}{2}\right)
=
\kappa 
\left(U-\frac{\kappa}{2}\right),
$$
where $U\sim \Normal(0,1)$.
Hence, this has a $\Normal\left(-\frac{1}{2}\kappa^2,\kappa^2\right)$ distribution. By Proposition 2.4 of \citet{roberts1997weak}, this leads to the corresponding limiting acceptance probability stated in the theorem.

\section{Proofs of ergodicity results}

\subsection{Proof of Proposition \ref{prop.isotropic.ergodic}}
\label{app.prove.ergodic.prop}
Since the algorithm targets $\pi$ by design, using Theorem 4 of \citet{RobRos2004}, we must show that the algorithm is $\phi$-irreducible and aperiodic.

Consider the first reflection step in the Hug proposal, starting at $x_0$, with a proposed velocity of $\lambda Z$ where $Z\sim \Normal(0,I_d)$. By symmetry the next point, $x_1$ is on the same (spherical) contour as $x_0$ ($||x_1||=||x_0||)$, and the unit gradient vector at the reflection point is $(x_0+x_1)/||x_0+x_1||$. The proposal is $X_1=x_1$ whenever $\lambda \delta Z=x_1-x_0+a (x_0+x_1)/||x_0+x_1||$ for any $a \in \mathbb{R}$. Since $Z$ is Gaussian, the $d-1$ dimensional density $f_{X_1|X_0}(x_1|x_0)$ with respect to Lebesgue measure on the hypersphere with $||x_1||=||x_0||$ is positive and continuous for all $x_1$ and $x_0$. By induction, therefore, the density for $X_B$, $f_{X_B|X_0}(x_B|x_0)$ on the same hypersphere is positive and continuous for all $x_B$.

Since $||x_B||=||x_0||$ and the proposal density is isotropic, the acceptance probability for the proposal is $1$. Thus $\Phug$ is encapsulated by the density $f_{X_B|X_0}(x_B|x_0)$ which is strictly positive and continuous across the whole hyperspherical surface.

Hence, the combination of the Hug and Hop kernels is
\[
P_{HH}=\int f_{X_B|X_0}(x_B|x_0) \Phop(x_B,\cdot)\md x_B.
\]
Thus, $P_{HH}$ can be viewed as similar to a Metropolis-Hastings kernel with an acceptance probability of $\alpha_{Hop}(x_B,y)$, except that even if a rejection occurs there is movement, from $x_0$ to $x_B$.

The proofs in \citet{RobRos2004} that the Running Example is both $\phi$-irreducible and Harris recurrent only use the consequences of an acceptance, so they apply equally well here. They also require that the density, here $f(||x||)$, is finite everywhere and the proposal is positive and continuous everywhere in $\mathbb{R}^d$ from any starting point in $\mathbb{R}^d$. We have the finiteness by assumption.

When $\mu\ne 0$, the Gaussian hop proposal has support over $\mathbb{R}^d$, whatever $x_B$ and we are done. Because the target is isotropic, the gradient at $x_B$ is $\pm x_B/||x_B||$; so, when $\mu=0$, Hop only proposes moves along the line that includes the origin and $x_B$. The proposed movement along this line is $\mathsf{N(0,\lambda^2/(1\vee||g(x)||^2)}$, which has support across the whole line. Since the combination of movement anywhere on the hyperspherical surface and then movement anywhere along the radial line corresponds to movement anywhere in $\mathbb{R}^d$, the proposal has support across $\mathbb{R}^d$, and is continuous because it is the convolution of two continuous functions.

\subsection{Proof of Theorem \ref{thm.HopGE}}
\label{app.GEHop}
From a current position $x\in \mathbb{R}$, the \Hop algorithm on a target of the form \eqref{eqn.oneDtarget} is a Metropolis-Hasting algorithm with a proposal density of
\begin{align}
  \label{eqn.oneDhopprop}
  q(y\mid x)
  &=
   \frac{s(x)}{\sqrt{2\pi \lambda^2}}
   \exp\left[-\frac{1}{2\lambda^2}s(x)^2(y-x)^2\right],~~~y\in\mathbb{R}.
\end{align}

The \Hop proposal in \eqref{eq:hop-prop} has, for targets of the form \eqref{eqn.oneDtarget}, $s(x)=||\nabla \log \pi(x)||=|x|^{a-1}$. But later in Section \ref{sec:hop} it is pointed out that this is degenerate when the gradient is zero (here at $x=0$). Hence the theorem uses $s(x)=1\vee ||\nabla \log \pi||=1\vee |x|^{a-1}$.

Firstly, when $a=1$, the algorithm is simply an RWM on a Laplace target and so is geometrically ergodic \citep[]{MengTwee1996}. So for the remainder of the proof we restrict attention to $a>1$.

To prove geometric ergodicity we will use the following standard result.
\begin{theorem}
  \citep[A slight simplification of Theorem 9 of][]{RobRos2004}
 Consider a $\phi$-irreducible aperiodic Markov chain with a kernel of $P$ and a stationary distribution of $\pi$ on a space $\cX$. Suppose the minorisation condition \eqref{eqn.minorisation} is satisfied for some $C\subset \cX$ and $\epsilon>0$ and probability measure $\nu$. Suppose further that the drift condition \eqref{eqn.genDrift} is satisfied for some constants $0<\beta<1$ and $b<\infty$, and a function $V:\cX\rightarrow [1,\infty]$ with $V(x)<\infty$ for at least one $x\in \cX$. Then the chain is geometrically ergodic.
 \begin{align}
  \label{eqn.minorisation}
  P(x,A)&\ge \epsilon \nu(A)~~~\forall~x\in C~\mbox{and measureable}~A\subseteq \cX,\\
  PV(x):= \int P(x,\md y)V(y)&\le \beta V(x)+b \mathbf{1}_C(x)~~~\forall~x\in\cX.
   \label{eqn.genDrift}
 \end{align}
\end{theorem}

From a current value $x\in \mathbb{R}$, the acceptance probability for a \Hop proposal $y\in\mathbb{R}$ is $1\wedge r(x,y)$, where $r(x,y)$ is the acceptance ratio:
\begin{align*}
  r(x,y)&
   :=
   \frac{\pi(y)q(x\mid y)}
   {\pi(x)q(y\mid x)}.
\end{align*}
Any Metropolis-Hastings chain where there is a chance of rejection is aperiodic, and because the proposal has positive support on the whole of $\cX$, combined with a positive acceptance probability, the \Hop algorithm is irreducible.
For the robust proposal \eqref{eqn.oneDhopprop}, consider sets of the form $C=[-c,c]$ for some $c\ge 1$. For $x\in C$, $(y-x)^2\le (|y|+c)^2$,
$\inf_{x \in C}q(y|x)\ge \frac{1}{\sqrt{2\pi\lambda^2}}\exp[-c^{2a-2}(|y|+c)^2/(2\lambda^2)]>0$ and $\inf_{x \in C}q(x|y)=\frac{1}{\sqrt{2\pi\lambda^2}}\sqrt{1\vee |y|^{2a-2}}\exp[-(1\vee |y|^{2a-2})(|y|+c)^2/(2\lambda^2)]>0$. Thus
\begin{align*}
  q(y|x)\alpha(x,y)&=
   q(y|x)\wedge \frac{\pi(y)q(x|y)}
   {\pi(x)}
   \ge
   \inf_{x\in C}q(y|x)\wedge
   \left[\frac{1}{\sup_{x\in C}\pi(x)}\pi(y)\inf_{x\in C}q(x|y)
   \right]>0,
 \end{align*}
showing that the minorisation condition \eqref{eqn.minorisation} is satisfied.  It remains to show that the drift condition \eqref{eqn.genDrift} is satisfied. For $x\in C$, $PV(x)=A(x)+B(x)$, where $A(x)=V(x) [1-\int_{-\infty}^\infty q(y|x)\alpha(x,y)~\md y] \le V(x)$, and
\begin{align*}
  B(x)
  &=
   \int_{-\infty}^\infty V(y)\alpha(x,y)q(y|x)\md y
   =\int_{-\infty}^\infty [q(y|x)V(y)]\wedge [q(x|y)V(x)^2/V(y)]\md y\\
  &<V(x)^2\int_{-\infty}^\infty q(x|y)/V(y)\md y
   \le V(c)^2\frac{1}{\sqrt{2\pi\lambda^2}}
   \int_{-\infty}^\infty (1\wedge |y|^{a-1})
   \exp\left[
  %     -\frac{1}{2\lambda^2}}(1\wedge |y|^{2a-2})(y-c)^2
   -\frac{1}{2a}y^{a}\right]\md y<\infty.
\end{align*}
 To complete the proof, we must show that for some $C=[-c,c]$,
$PV(x) /V(x)\le \beta$ for all $x\notin C$.

For any current value $x \in \mathbb{R}$, we
define the acceptance region, $\Accept(x):=\{y: r(x,y)\ge 1\}$ and let $\Reject(x):=\Accept(x)^c$, be the region where rejection is possible.

\begin{align}
  \nonumber
  \frac{PV(x)}{V(x)}
  &=
   \frac{1}{V(x)}\int_{\cX}V(y) \alpha(x,y) q(y|x) \md y
   +
   \left[1-\int_\cX \alpha(x,y)q(y|x)\md y\right]\\
  \label{eqn.drift.form.one}
  &=
   1
   +
   \frac{1}{V(x)}\int_{\cX}[V(y)-V(x)] \alpha(x,y) q(y|x) \md y\\
  &=
   1
   +\int_{\Accept(x)} \left[\frac{V(y)}{V(x)}-1\right]q(y|x)\md y
   +\frac{1}{V(x)}\int_{\Reject{x}} [V(y)-V(x)] \frac{\pi(y)}{\pi(x)}q(x|y)\md y.
   \label{eqn.drift.form.two}
\end{align}
As with some proofs of the geometric ergodicity of the RWM \citep[e.g.][]{RobTwee1996RWM}, we take $V(x):=1/\sqrt{\pi(x)}=\exp[|x|^a/(2a)]$. By symmetry it is sufficient to consider the behaviour for positive $x$. We first show that if $c$ is large enough the acceptance region is that same as for the RWM.

\begin{lemma}
  For the \Hop proposal \eqref{eqn.oneDhopprop} on a target of the form
 \eqref{eqn.oneDtarget}, for every $a>1$ there is a finite $c_*(a)>0$ such that for all $x$ with $|x|\ge c_*(a)$, $\Accept(x)=\{y:|y|\le |x|\}$.
\end{lemma}

\begin{proof}
  Firstly, define $h_a(x):=\frac{1}{a}|x|^a - \frac{1}{2}\log(1+|x|^{2a-2})$
 and $g_a(x,y)=\frac{1}{2}(|x|^{2a-2}-|y|^{2a-2})(y-x)^2$. Then
 \begin{align*}
  \Accept(x)&
   :=
   \left\{
   y: h_a(x)-h_a(y)+\frac{1}{\lambda^2}g(x,y)\ge 0
   \right\}.
 \end{align*}
 From the form of $h_a$, there exists some $x_1=c_1(a)<\infty$ such that $h_a(x)$ is monotonically increasing in $|x|$ for all $|x|>x_1(a)$. Also $h_a(0)=0$ and $h_a$ is continuous, so $h_a$ has a finite upper bound on $[-x_1,x_1]$ which we denote $h_a^*:=\sup_{x\in[-x_1,x_1]}h_a(x)$. Since $h_a$ increases without bound as $|x|\uparrow \infty$, $c_*:=\inf\{x\ge x_1:h_a(x)\ge h_a^*\}$ is well defined. For any $x$ with $|x|\ge x_1$, and any $y$, we have, therefore that if $|y| \le |x|$ then $h_a(y)\le h_a(x)$ and if $|y|>|x|$ then $h_a(y)>h_a(x)$. Finally, $|y|\le |x|\Leftrightarrow g_a(x,y)\ge 0$, so $|y|\le |x| \Rightarrow h_a(x)-h_a(y)+g_a(x,y) \ge 0$ and $|y|>|x|\Rightarrow h_a(x)-h_a(y)+g_a(x,y)<0$, as required.
\end{proof}

Conditional on $x\ge c_*$, we next partition $\Reject(x)$ into three regions: $C_1:=(-\infty,-x)$, $C_4:=[x,x+x^{2-a})$
 and $C_5:=[x+x^{2-a},\infty)$, and we partition $\cA(x)$ into $C_2:=[-x,x-x^{2-a})$ and $C_3:=[x-x^{2-a},x)$.

For integrands within $\Reject(x)$, and with $V(x)=1/\sqrt{\pi(x)}$, $V(x)\le V(y)$,
 we will use the following trivial equivalence.
\begin{proposition}
  \label{prop.Vmanip}
 With $V(x)=1/\sqrt{\pi(x)}$ and $\pi(y)\le \pi(x)$,
 \[
  [V(y)-V(x)]\frac{\pi(y)}{\pi(x)}
 =
 V(x)\left[1-\frac{V(x)}{V(y)}\right]\frac{V(x)}{V(y)}.
\]
\end{proposition}

We write $s(x)=|x|^{a-1}$ rather than $s(x)=1\vee|x|^{a-1}$ because: (i) for $q(y|x)$ we have $|x|\ge c$, and we may choose $c\ge 1$, and (ii) $q(x|y)$ is only every required for $y\in \Reject(x)$, for which $|y|\ge |x|\ge c\ge 1$.

We now show that the contribution to $PV/V$ from regions $C_1$, $C_2$ and $C_5$ can be made negligible.

\begin{lemma}
  \label{lem.negRegs}
 \begin{align*}
  \frac{1}{V(x)} \int_{C_1\cup C_2 \cup C_5}[V(y)-V(x)]\alpha(x,y)q(y|x)\md y
  &<
   \frac{4}{\sqrt{2\pi\lambda^2}}
   \exp\left[-\frac{1}{2\lambda^2}x^{2}\right].
 \end{align*}
\end{lemma}

\begin{proof}
  From Proposition \ref{prop.Vmanip}, the integrand in regions $\subseteq \Reject(x)$ in \eqref{eqn.drift.form.two} can be rewritten as
\begin{align*}
  f_{\cR}(y;x,a,\lambda)&
   :=
   \frac{V(x)}{V(y)}\left[1-\frac{V(x)}{V(y)}\right]
   \frac{|y|^{a-1}}{\sqrt{2\pi\lambda^2}}
   \exp\left[
   -\frac{1}{2\lambda^2}|y|^{2a-2}(y-x)^2
   \right].
 \end{align*}
 In $C_1\cup C_5$, $|y|\ge x$ and $|y-x|\ge x^{2-a}$, so
 $|y|^{2a-2}(y-x)^2\ge x^2$.
Thus
 \begin{align*}
  f_{\cR}(y;x,a,\lambda)
  &\le
   \frac{V(x)}{V(y)}\frac{|y|^{a-1}}{\sqrt{2\pi\lambda^2}}\exp\left[-\frac{1}{2\lambda^2}x^2\right]\\
  &=
   \exp\left[\frac{1}{2a}x^a\right]\frac{1}{\sqrt{2\pi\lambda^2}}
   \exp\left[-\frac{1}{2\lambda^2}x^{2}\right]
   \times |y|^{a-1}\exp\left[-\frac{1}{2a}|y|^a\right].
 \end{align*}
 So
 \begin{align*}
  \int_{-\infty}^{-x}f_{\cR}(y;x,a,\lambda)\md y
  &\le
   \exp\left[\frac{1}{2a}x^a\right]\frac{1}{\sqrt{2\pi\lambda^2}}
   \exp\left[-\frac{1}{2\lambda^2}x^{2}\right]
   \times
   2\exp\left[-\frac{1}{2a}x^a\right]\\
  &=
   \frac{2}{\sqrt{2\pi\lambda^2}}
   \exp\left[-\frac{1}{2\lambda^2}x^{2}\right].
 \end{align*}
 Since the bound on $f_{\cR}(y;x,a,\lambda)$ is positive and is symmetric in its first argument,\\ $\int_{C_5} f_{\cR}(y;x,a,\lambda)\md y<\int_{C_1} f_{\cR}(y;x,a,\lambda)\md y$. Finally, in $C_2$, $V(y)\le V(x)$ so
 $\int_{C_2}[V(y)-V(x)]q(y|x)\md y<0$. Combining the three inequalities gives the required result.
\end{proof}

It remains to consider the integrals over $C_3$ and $C_4$. We now provide a simplification of the integral over $C_4$. Define
\begin{align*}
  D(x)&:=
   \frac{1}{V(x)}\int_{C_4}[V(y)-V(x)]\alpha(x,y)q(y|x)\md y
   -\frac{1}{V(x)}\int_{C_4}[V(y)-V(x)]\frac{\pi(y)}{\pi(x)}q(y|x)\md y.
 \end{align*}

\begin{lemma}
  \label{lem.qswap}
 \[
  D(x)<\frac{1}{2}\left[(1+x^{1-a})-1\right].
\]
\end{lemma}

\begin{proof}
  Over the range of the integrand, since $y\ge x$, and $y<x+x^{2-a}=x(1+x^{1-a})$,
 \begin{align*}
  \frac{q(x|y)}{q(y|x)}
  &=
   \frac{y^{a-1}}{x^{a-1}}
   \exp\left[
   -\frac{1}{2}\left(y^{2a-2}-x^{2a-2}\right)(y-x)^2
   \right]\\
  &<\frac{y^{a-1}}{x^{a-1}}=\left(1+x^{1-a}\right)^{a-1}.
 \end{align*}
Thus,
 \begin{align*}
  D(x)
  &=
   \int_{C_4}\frac{V(x)}{V(y)}\left[1-\frac{V(x)}{V(y)}\right]
   \left[\frac{q(x|y)}{q(y|x)}-1\right]q(y|x)\md y\\
  &<
   \left[(1+x^{1-a})-1\right]
   \int_{x}^{\infty} q(y|x)\md y,
 \end{align*}
 giving the required result, since the integral is $1/2$.
\end{proof}

Lemma \ref{lem.negRegs} tells us that the contribution to $PV(x)/V(x)$ from regions outside of $[x-x^{2-a},x+x^{2-a})$ can be made as small as desired by taking $x$ sufficiently large. Since $a>1$. Lemma \ref{lem.qswap} tells us that the positive upper bound on the discrepancy from integrating with respect to $q(y|x)\md y$ rather than $[q(x|y)/q(y|x)]q(y|x)\md y$ can be made negligible. Thus it remains to show that
\[
  T(x):=\int_{C_3}\left[\frac{V(y)}{V(x)}-1\right]q(y|x)\md y
+
\int_{C_4}\frac{V(x)}{V(y)}\left[1-\frac{V(x)}{V(y)}\right]q(y|x) \md y
\]
is strictly negative.

\begin{lemma}
  For any $\epsilon>0$, there exists $x(\epsilon)$ such that for any $x>x(\epsilon)$
 \[
  T(x)< - \int_0^{x/\lambda}(1-\exp[-\lambda z])^2\phi(z)\md z + \epsilon,
\]
 which can be made strictly negative by taking $x$ sufficiently large.
\end{lemma}

\begin{proof}
  Set $Y=x+\lambda Z/x^{a-1}$, where $Y$ has the density $q(y|x)$, so that $Z\sim \Normal(0,1)$, and denote its density function by $\phi(z)$. Then, since $y=x\pm x^{2-a}\Rightarrow z=\pm x/\lambda$,
 \begin{align*}
  T(x)&=
   \int_{-x/\lambda}^0
   \left[\frac{V(x+\lambda z/x^{a-1})}{V(x)}-1\right]\phi(z)\md z
   +
   \int_0^{x/\lambda}
   \frac{V(x)}{V(x+\lambda z/x^{a-1})}\left[1-\frac{V(x)}{V(x+\lambda z/x^{a-1})}\right] \phi(z)\md z\\
  &=
   \int_0^{x/\lambda}
   \left\{\frac{V(x-\lambda z/x^{a-1})}{V(x)}-1+\frac{V(x)}{V(x+\lambda z/x^{a-1})}-\frac{V(x)^2}{V(x+\lambda z/x^{a-1})^2}\right\} \phi(z)\md z,
 \end{align*}
 as $\phi$ is an even function.
 However
 \begin{align*}
  \log\left[\frac{V(x-\lambda z/x^{a-1})}{V(x)}\right]
  &=
   \frac{1}{2a}\left\{(x-\lambda z/x^{a-1})^{a}-x^a\right\}\\
  &=
   \frac{x^a}{2a}\left\{(1-\lambda z/x^{a})^a-1\right\}\\
  &=
   \frac{x^a}{2a}\left\{-a\lambda z/x^a+\mathcal{O}((\lambda z/x^a)^2)\right\}\\
  &=
   -\lambda z/2 + \mathcal{O}\left((\lambda z)^2/x^a\right).
 \end{align*}
 Similarly
 \begin{align*}
  \log\left[\frac{V(x)}{V(x+\lambda z/x^{a-1})}\right]
  &=
   -\lambda z/2 + \mathcal{O}\left((\lambda z)^2/x^a\right).
 \end{align*}
 So, as $x\rightarrow \infty$,
 \begin{align*}
  \frac{V(x-\lambda z/x^{a-1})}{V(x)}-\frac{V(x)}{V(x+\lambda z/x^{a-1})}
  &=
   \exp(-\lambda z)
   \left\{
   \exp\left[\mathcal{O}\left(\frac{\lambda^2 z^2}{x^a}\right)\right]
   -
   \exp\left[\mathcal{O}\left(\frac{\lambda^2 z^2}{x^a}\right)\right]
   \right\}
   \rightarrow 0
 \end{align*}
 for any fixed $z$. Thus, Since both
 $0\le \frac{V(x-\lambda z/x^{a-1})}{V(x)}\le 1$ and
 $0\le \frac{V(x)}{V(x+\lambda z/x^{a-1})}\le 1$,
\begin{align*}
  \int_0^{x/\lambda}
  \left[\frac{V(x-\lambda z/x^{a-1})}{V(x)}-\frac{V(x)}{V(x+\lambda z/x^{a-1})}\right]\phi(z)\md z
  \rightarrow 0,
 \end{align*}
as $x\rightarrow \infty$ by the Dominated convergence Theorem, and we may instead consider
 \begin{align*}
  T_*(x)
  &=
   \int_0^{x/\lambda}
   \left\{\frac{V(x)}{V(x+\lambda z/x^{a-1})}-1+\frac{V(x)}{V(x+\lambda z/x^{a-1})}-\frac{V(x)^2}{V(x+\lambda z/x^{a-1})^2}\right\} \phi(z)\md z\\
  &=
   - \int_0^{x/\lambda}
   \left[1-\frac{V(x)}{V(x+\lambda z/x^{a-1})}\right]^2\phi(z) \md z.
 \end{align*}
 The integrand is bounded above by $\phi(z)$, so by the Dominated Convergence Theorem, as $x\rightarrow \infty$,
 \begin{align*}
  T_*(x)&\rightarrow
   \int_0^\infty[1-\exp(-\lambda z/2)]^2\phi(z).
 \end{align*}
\end{proof}

\section{Empirical investigation of convergence and efficiency of Hop}
\label{sec.emp.Hop}
\begin{figure}[ht!]
%\begin{figure}[]
  \centering
  \includegraphics[height=0.5\textheight,width=0.75\textwidth]{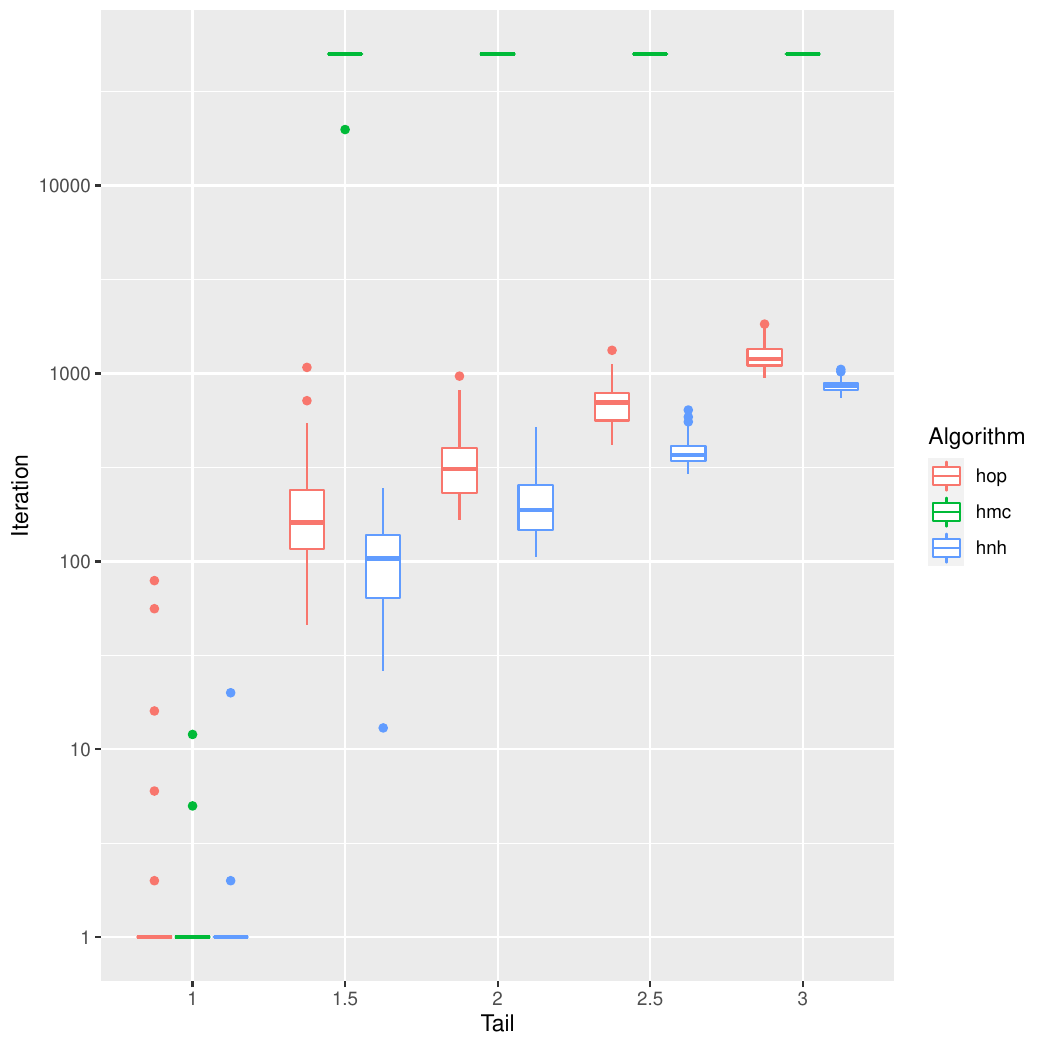}
  \caption{Iteration number at which Hop, HMC, and Hug and Hop converged to the
    main mass of the target ~(\ref{eq:expnormtarget}) with
    $a=4$. The x-axis denotes the starting multiplier $\gamma$.}
  \label{fig:hop_hmc_tail_iter}
\end{figure}

Figure \ref{fig:hop_hmc_tail_iter} shows the results from the empirical study described in Section \ref{sec.ergANDconv}.

\section{Example targets}
\label{app:example-targ-calc}

Firstly, consider an equal mixture of two $\Normal\left(\mu_i, \Sigma_i\right)$ distributions,
then $\EE{X} = \half(\mu_1+\mu_2)$ and
\begin{align}\label{eq:mix_cov}
  \Cov{X} &= \half \Sigma_1 + \half \Sigma_2
            + \recip{4}(\mu_1-\mu_2)(\mu_1-\mu_2)^\top.
\end{align}

\subsection{Banana}
The Banana target is parameterised by $\lambda \in (0,1)$, its bananacity.
The two components $(x_1,x_2)$ satisfy:
\[
  X_1 \sim\Normal(0,1), \qquad X_2|x_1 \sim \Normal(\lambda (x_1^2-1), 1-\lambda^2).
\]
Values
of $\lambda$ closer to one make the
banana bendier, whilst at $\lambda=0$ the target degenerates to a $\Normal\left(0,I\right)$.
The log-target for this model is thus:
\[
  \log\pi(x) = -\frac{x_1^2}{2} - \frac{(x_2 - \lambda (x_1^2 - 1))^2}{2(1-\lambda^2)}.
\]

\subsection{Bimodal}
\label{sec.BimodalDetail}
The Bimodal is an equal mixture of two bivariate Normal
distributions:
\[
  X \sim \half \Normal(-\mu, \Sigma) + \half \Normal(\mu, \Sigma).
\]
with $\mu = \sqrt{\lambda}\bone$ and $\Sigma = (1-\lambda) I$. Thus, $\Expect{X}=0$ and, by (\ref{eq:mix_cov}),
$\Cov{X} = \Sigma + \mu\mu^\top = (1-\lambda)I + \lambda \bone \bone^\top$.
For the main experiments, the results of which are summarised in Figure \ref{fig:sim-ess-per-sec}, we set $\lambda=0.95$.

For the extreme experiment at the end of Section \ref{sec:sims} we set $\lambda=0.9$ and an overall scale for component $i$ of $1+9(d-i)/(d-1)$. We chose these values so that the algorithms with no preconditioning were able to travel between the modes, but that such movement happened relatively rarely.

Over five replicate experiments, each of $2\times 10^5$ iterations, the best-performing NUTS algorithm used $\epsilon=1.95$, which led to an acceptance rate of $84\%$, a mean CPU time of $134\%$
% actually $269\%$ but am dividing by 2 as a more cunning implementation
% would only use one gradient calculation per leapfrog
of that of HMC and $(17,14,10,23,16)$ mode flips (mean$=16$). The best performing HMC algorithm used $T=20$ and $L=15$, which led to an acceptance rate of $83\%$ and $(31,28,37,34,36)$ mode flips (mean$=33.2$). The best performing Hug and Hop algorithm used $T=30$, $B=16$, $\lambda=5$ and $\kappa=1$, which led to acceptance rates of $\alpha_{hug}=63\%$ and $\alpha_{hop}=23\%$, a mean CPU time of $149\%$ of that of HMC and $(75,83,67,63,77)$ mode flips (mean$=73$).
% Timings: NUTS: 65.3 (but should nearly half it), HMC: 24.3, HnH: 36.1

\begin{figure}[ht!]
  \centering
  \includegraphics[width=0.8\textwidth]{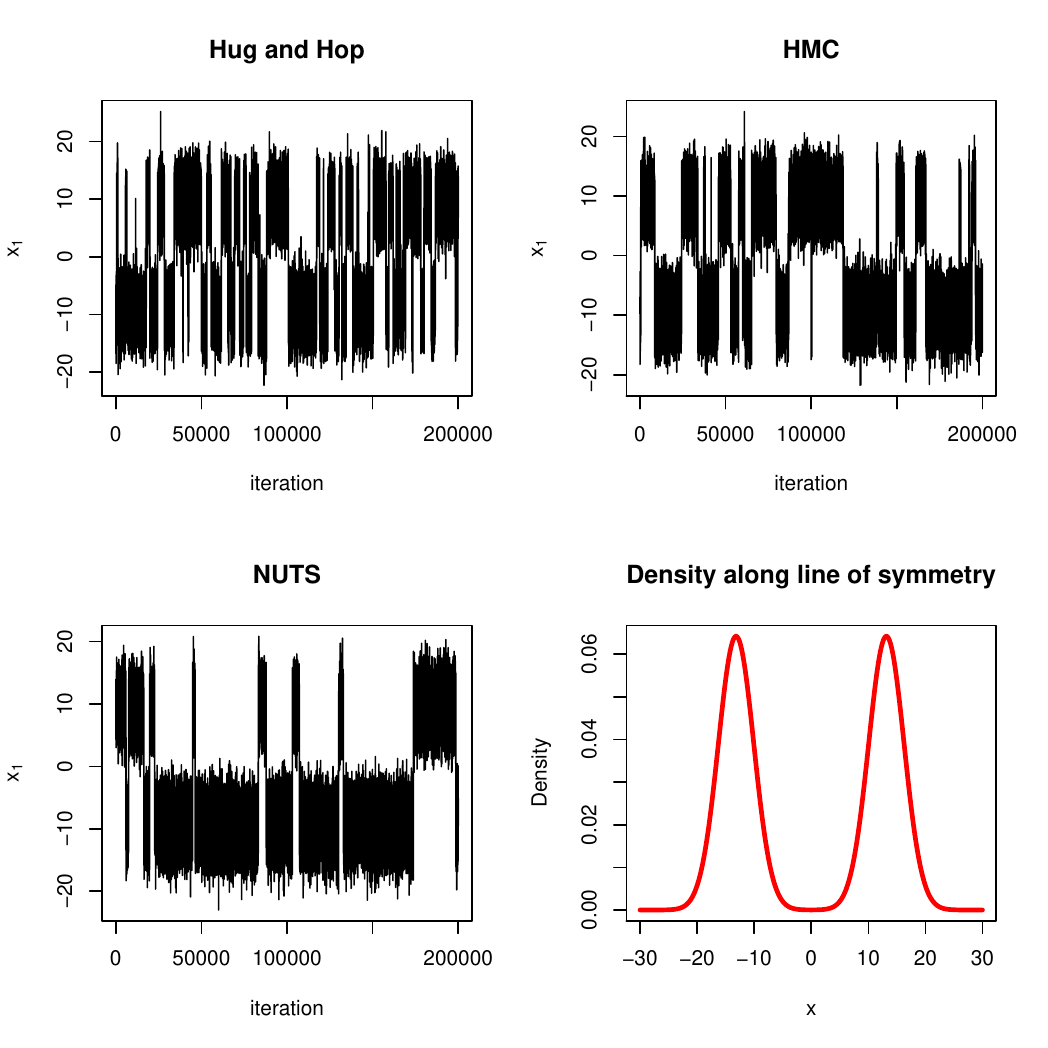}
  \caption{Trace plots for $x_1$ from Hug and Hop, HMC and NUTS for the extreme bimodal target, and the (both marginal and conditional) density along the line of symmetry between the modes.
  \label{fig:bimodal_extreme}}
\end{figure}

Figure \ref{fig:bimodal_extreme} provides trace plots for each of the three algorithms from the final replicate for each experiment, as well as the true density along the line between the two modes. The scale of $x_{25}$, which is $1$, restricts the sizes of the steps that lead to reasonable acceptance rates; the size of the gap between the modes should be viewed relative to this. The algorithms were tuned so to maximise mode hopping, nonetheless, when comparing the minimum effective sample size over the $23$ unimodal components, that of Hug and Hop was $163\%$ of that of HMC, indicating that it is slightly more efficient in these terms, too.  

\subsection{PlusPrism}
The PlusPrism is an equal mixture of two centred
bi-variate Normal distributions with covariance matrices $\Sigma_1=
\diag(1+\lambda, 1-\lambda)$ and $\Sigma_2= \diag(1-\lambda, 1+\lambda)$.
The overall mean is at $\bzero$, while the covariance is given by
$\Cov{X} = (\Sigma_1+\Sigma_2)/2 = I$. We set $\lambda=0.95$.

This target has mass spread in a ``+'' shape along the $x$ and $y$
axis with a mode at (0,0).  In three or more dimensions, this two-dimensional
plus is projected along the other dimensions creating a prism.

\section{Statistical models}
\subsection{Cauchit regression}
\label{app:cr-calc}
To simplify the formulae, we redefine the response to be $Y_i\in \{-1, 1\}$ rather than $Y_i\in \{0,1\}$. The inverse link function is
$g^{-1}(x) = 1/2 + \arctan(x)/\pi$,
where, here only, $\pi$ is the number $3.14\dots$.
Now, $g^{-1}(-x) = 1 - g^{-1}(x)$, and writing
$\eta_i = x_i^\top \beta$,
%   and the log-posterior
%     $\log\pi(\beta| y, x) = \ell$ and its gradient are:
\begin{align*}
  \ell(\beta) &= -\frac{\tau}{2}\norm{\beta}^2 + \sum_{i} \log \left(1/2 + \arctan(y_i\eta_i)/\pi\right),\\
  \partfrac{\ell}{\beta_j} &= -\tau\beta_j + \sum_{i} \frac{y_i x_{ij}}{(1+\eta_i^2)\left(\pi/2 + \arctan(y_i\eta_i)\right)}.
\end{align*}

\begin{figure}[ht!]
  \centering
  \includegraphics[width=0.8\textwidth]{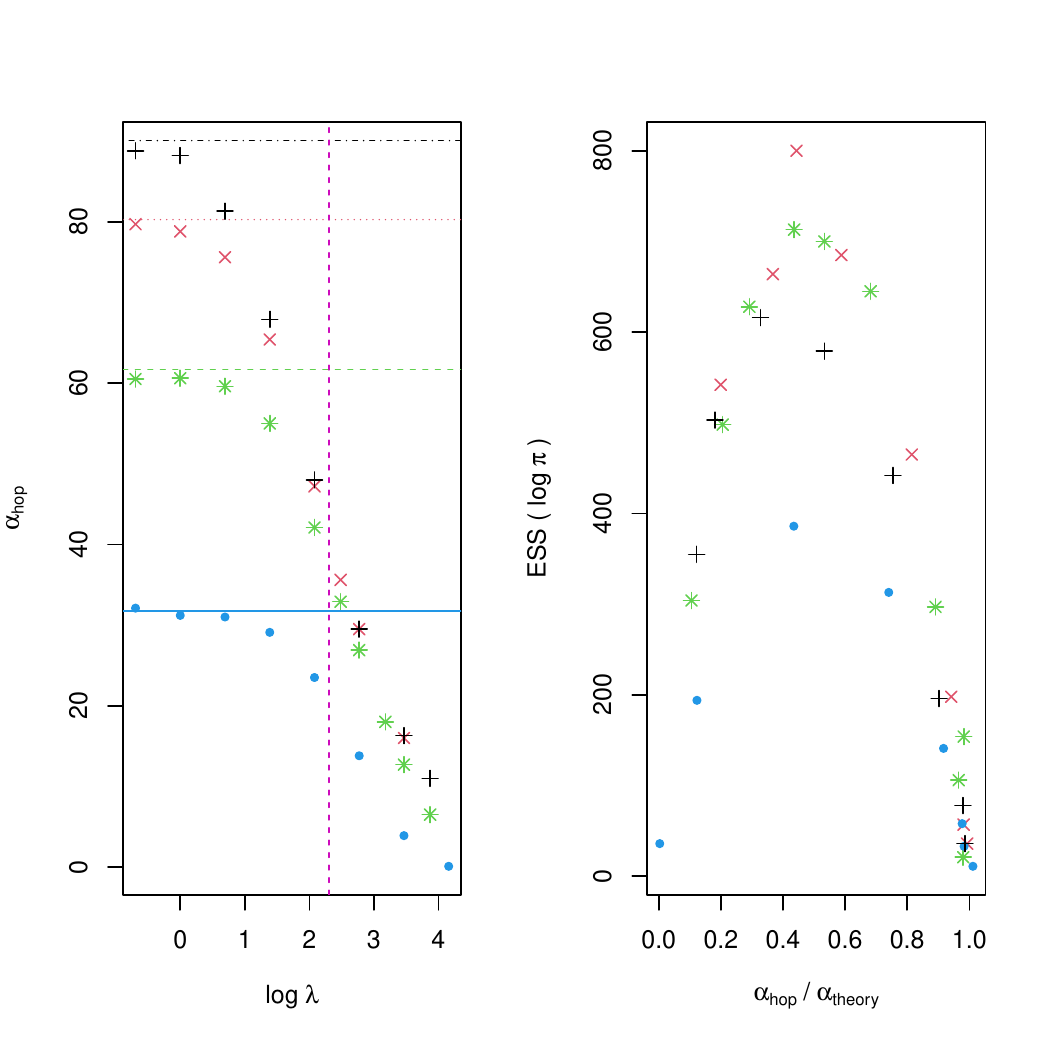}
  \caption{Hug and Hop applied to a 100-dimensional Cauchit-regression posterior with Hug parameters fixed at $(T=0.32,B=4)$. Hop parameters were varied, with and $\lambda \in\{1/2,1,2,4,8,12,16,32,48\}$ and $\kappa\in\{1/4,1/2,1,2\}$. Points and lines are coloured according to $\kappa$: black $+$ and dot-dashed-line ($\kappa=1/4$); red $\times$ and dotted line ($\kappa=1/2$); green $*$ and dashed line ($\kappa=1$); blue $\bullet$ and solid line ($\kappa=2$). The left plot shows the acceptance rate as a function of $\lambda$, with horizontal lines marking the theoretical acceptance rate, the vertical magenta line corresponds to $\lambda=d^{1/2}=10$; the right plot shows the effective sample size of $\log \pi$ as a function of the ratio of the observed acceptance rate to the asymptotic acceptance rate.
  \label{fig:cauchit_explore}}
\end{figure}

Figure \ref{fig:cauchit_explore} shows the effect of varying the \emph{Hop} tuning parameters on the \emph{Hop} acceptance rate and efficiency of exploration of the 100-dimensional Cauchit-regression posterior. The \emph{Hug} tuning parameters were set to values that explored the posterior adequately, but not optimally; very similar patterns were found with other settings for the Hug parameters where mixing was at least adequate. 

The left-hand plot shows that for small to moderate values of $\lambda$, the acceptance rate is close to the theoretical value, but as $\lambda$ increases towards and then beyond $d^{1/2}=10$ the acceptance probability drops monotonically towards zero. The right-hand plot shows that in this example, whatever the setting of $\kappa$, the optimal choice of $\lambda$ is achieved when the acceptance rate is around a third to a half of the asymptotic rate for that $\kappa$ value; \emph{i.e.}, when the asymptotics have started to break down but have not completely broken down.

\subsection{Rasch model}
\label{app:irm-calc}
As with the Cauchit regression we redefine the response to be $Y_{ij} \in \{-1,1\}$ and let $z_{ij}= y_{ij}(\eta_i - \beta_j)$. Then:
\begin{align*}
  \ell(\beta, \eta | y) &= \sum_{i,j=1}^{M,N} \log\Phi(z_{ij}) -\frac \tau 2 \sum_{i=1}^M \eta_i^2- \frac \tau 2 \sum_{j=1}^M\beta_j^2,\\
  \partfrac{\ell}{\eta_k} &= \sum_{j=1}^N y_{kj}\frac{\phi(z_{kj})}{\Phi(z_{kj})} - \tau\eta_k,\\
  \partfrac{\ell}{\beta_k} &= \sum_{i=1}^M -y_{ik}\frac{\phi(z_{ik})}{\Phi(z_{ik})} - \tau\beta_k.
\end{align*}

% \subsection{Probit Spatial Regression}
% \label{app:psr-calc}
% Recall the model in Equation~(\ref{eq:PSR}) and the reparameterisation
% into ($Z, \btheta$). Let $[Av]_i$ denote the $ith$ element of vector
% resulting from the matrix product $Av$ and $\Sigma = A^TA$. Then $\Var{X} =
% \Var{A^T Z} = \Sigma$
% With $Y_{g} \in \{-1,1\}$, we obtain as the log-target:
% \[
%   \ell(\rho, \psi, z | y) = \sum_g \log\Phi(y_g[A^\top z]_g) - \frac 1 2 z^\top z - \frac 1 2 \rho^2 - \frac 1 2 \psi^2,
% \]
% and the gradient with respect to a component of $z$ is
% \begin{align*}
%   \partfrac{\ell}{z_j} &= -z_j + \sum_g y_g A^\top_{gj} \frac{\phi(y_g[A^\top z]_g)}{\Phi(y_g[A^\top z]_g)}.
% \end{align*}
%   which, with $\cdot$ as element wise multiplication, yields:
%     \begin{align*}
  %     \nabla_z\ell &= -z + A \left[y\cdot\frac{\phi(y \cdot [A^\top z])}{\Phi(y \cdo%t [A^\top z])}\right].
   %   \end{align*}

\subsection{Stochastic volatility model}\label{app:stochvol}
Let $t = 0,\dots, T-1$ be $T$ index equally spaced moments in time and consider the following model:
\[\begin{aligned}
z_t &\sim \Normal(0, 1)~~~\mbox{i.i.d},\\
x_0 &= \frac{z_0}{\phi},\\
x_t &= \rho x_{t-1} + z_t,\\
y_t &\sim \Normal\left(0, \frac{\exp(2x_t)}{\tau}\right),
\end{aligned}\]
with $\phi=\sqrt{1-\rho^2}$ and prior distributions of $\tau\sim \GammaDist(21,5)$ and $(1+\rho)/2\sim \BetaDist(20,2)$.

\begin{figure}[ht!]
  \centering
  \includegraphics[page=3]{sv_data}
  \caption{Simulated data for the Stochastic volatility data study.}
  \label{fig:sv-data}
\end{figure}

The data simulated from this model and used in Section~\ref{sec:stochvol} are shown in Figure \ref{fig:sv-data}. To perform inference, we consider the posterior distribution on $z$ and the parameters $(\alpha, \beta)$, which map $(\tau, \rho)$ to the real line via the equations:

\[
\alpha = -\half \log(\tau)
, \quad
\beta  = \half\log\left(\frac{1+\rho}{1-\rho}\right)
\]
with inverses:
\[
\tau = \exp(-2\alpha)
, \quad
\rho = \frac{\exp(2\beta)-1}{\exp(2\beta)+1}=\tanh \beta.
\]
Up to additive constants, the prior log-densities for $\tau$ and $\rho$ are:
\[\begin{aligned}
\log f_\tau(\tau)
&= 20 \log(\tau) - 5\tau\\
\log f_{\rho}(\rho)
&=19\log(1+\rho)+\log(1-\rho)=20\log(1+\rho)-\log\left(\frac{1+\rho}{1-\rho}\right).
\end{aligned}\]

Thus, ignoring additive constants, the log-prior for $\alpha$ is:
\[
\begin{aligned}
\log f_\alpha(\alpha)
&= \log f_\tau(\tau(\alpha)) + \log \abspipes{\dbyd{\tau}{\alpha}}
= -40\alpha - 5\exp(-2\alpha) -2\alpha\\
&= -42 \alpha - 5\exp(-2\alpha).
\end{aligned}
\]
To obtain the log prior for $\beta$, $\log f_\beta(\beta)$, first note that:
\[
\abspipes{\dbyd{\rho}{\beta}}=\mathsf{sech}^2\beta=\frac{4 \exp(-2\beta)}{(1+\exp(-2\beta))^2},
\]
and
\[
\log(1+\rho)=\log\left(\frac{\exp(2\beta)}{\exp(2\beta)+1}\right)=-\log\left\{1+\exp(-2\beta)\right\}.
\]
Thus, ignoring additive constants,
\[\begin{aligned}
\log f_\beta(\beta)
&= \log f_\nu(\rho(\beta)) + \log \abspipes{\dbyd{\rho}{\beta}}\\
&= -20 \log(1+\exp(-2\beta)) -2\beta -2\beta-2\log\left\{1+\exp(-2\beta)\right\}\\
&= -22 \log\{1+\exp(-2\beta)\} -4\beta.
\end{aligned}\]

We now derive the log-posterior distribution in terms of $\alpha$ and $\beta$.
Since $\phi=\sqrt{1-\rho^2}=\mathsf{sech} {\beta}$, the model for the data is now:
\[\begin{aligned}
x_0 &= z_0 \cosh \beta,\\
x_t &= \tanh(\beta) x_{t-1} + z_t,\\
y_t &\sim \Normal\left(0, \exp(2\alpha +2x_t)\right).
\end{aligned}\]

Ignoring additive constants, the log-likelihood is
\[
\begin{aligned}
\ell(z,y; \alpha, \beta)
&=
- \half  \sum_{t=0}^{T-1} z_t^2
- \half \sum_{t=0}^{T-1}\left\{ 2\alpha +2x_t + \left(\exp(2\alpha+2 x_t)\right)^{-1}y_t^2\right\}.
\\
&=
-T\alpha
- \half  \sum_{t=0}^{T-1} z_t^2+2x_t+\exp(-2\alpha - 2x_t) y_t^2.
\end{aligned}
\]
Setting $w_t = \exp(-2x_t-2\alpha) y_t^2$, 
\[
\log\pi(z, \alpha, \beta | y)
=
-42 \alpha - 5e^{-2\alpha}-22\log\left(1+e^{-2\beta}\right) - 4\beta-T\alpha
-\sum_{t=0}^T x_t -\half \sum_{t=0}^{T-1} \left(w_t + z_t^2\right).
\]

\subsection{Gradients}
We have $\partial \log f_\alpha/\partial \alpha = -42+10\exp(-2\alpha)$ and
since $x_t$ does not depend on $\alpha$,
$\partial \ell/\partial \alpha = -T+\exp(-2\alpha)\sum_{t=0}^{T-1}\exp(-2x_t)y_t^2$. Thus
\[
\frac{\partial \log \pi}{\partial \alpha}
=
-42-T+10\exp(-2\alpha)+\exp(-2\alpha)\sum_{t=0}^{T-1}\exp(-2x_t)y_t^2.
\]
Now
\[
\frac{\partial \log f_\beta}{\partial \beta}
=
-22\left(\frac{-2 \exp(-2\beta)}{1+\exp(-2\beta)}\right)-4
=\frac{44}{1+\exp(2\beta)}-4.
\]
Also
\[\begin{aligned}
\frac{\partial \ell}{\partial \beta}
=
-\frac{1}{2}\sum_{t=0}^{T-1}2\frac{\partial x_t}{\partial \beta}
-2 \frac{\partial x_t}{\partial \beta}\exp(-2\alpha -2x_t)y_t^2\\
=
\sum_{t=0}^{T-1}\left\{\exp(-2\alpha -2x_t)y_t^2 -1\right\} \frac{\partial x_t}{\partial \beta}.
\end{aligned}\]
So
\[
\frac{\partial \log \pi}{\partial \beta}
=
\frac{44}{1+\exp(2\beta)}-4
+
\sum_{t=0}^{T-1}\left\{\exp(-2\alpha -2x_t)y_t^2 -1\right\} \frac{\partial x_t}{\partial \beta}.
\]
At $t=0$:
\[\begin{aligned}
\pbyp{x_0}{\beta} &= \pbyp{}{\beta}\left(z_0 \cosh \beta\right) = z_0 \sinh \beta
= \tanh(\beta) x_0
\end{aligned}\]

For $t=1 \to T-1$, $x_t = \tanh(\beta) x_{t-1} + z_t$, so
\[\begin{aligned}
\pbyp{x_t}{\beta} &=
 \tanh(\beta) \pbyp{x_{t-1}}{\beta} + \mathsf{sech}^2(\beta) x_{t-1}\\
\end{aligned}\]

We compute these terms recursively.

Finally,
\[\begin{aligned}
\frac{\partial \log \pi}{\partial z_s}&=
\frac{\partial \ell}{\partial z_s}
=
-
\sum_{t=0}^{T-1}z_t\II{t=s}+\frac{\partial x_t}{\partial z_s}
-2\frac{\partial x_t}{\partial z_s}\exp(-2\alpha-2x_t)y_t^2\\
&=
-z_s
+
\sum_{t=0}^{T-1}\left\{\exp(-2\alpha-2x_t)y_t^2-1\right\}\frac{\partial x_t}{\partial z_s}.
\end{aligned}\]

When $t=0$ we have:
\[
\pbyp{x_0}{z_s} = \cosh(\beta) \pbyp{z_0}{z_s} = \cosh(\beta) \II{s=0}.
\]

For $t=1 \to T-1$:
\[
\pbyp{x_t}{z_s}
= \pbyp{}{z_s}\left\{\tanh(\beta) x_{t-1} + z_t\right\}
= \tanh(\beta) \pbyp{x_{t-1}}{z_s} + \II{s=t}
\]

The solution to these recursions is:
\[
\pbyp{x_t}{z_s} = \tanh^{t-s}\beta\cosh^{\II{s=0}}\beta\II{s \leq t}.
\]
\end{document}